\documentclass[twocolumn,english,superscriptaddress,floatfix,longbibliography]{revtex4-1}
\usepackage[T1]{fontenc}
\usepackage[latin9]{inputenc}
\usepackage{textcomp}
\usepackage{amsmath}
\usepackage{amssymb}
\usepackage[pdftex]{graphicx}
\usepackage{xcolor}         

\makeatletter

\usepackage{braket}

\makeatother

\usepackage{babel}

\begin{document}

\title{Orbital Hall Effect and Angular Momentum Dynamics in Confined Geometries}
\author{Egor I. Kiselev}
\affiliation{Max Planck Institute for the Physics of Complex Systems, N\"othnitzer
Str. 38, 01187 Dresden, Germany}

\author{Benoit Doucot}
\affiliation{Laboratoire de Physique Th\'eorique et Hautes Energies,
Sorbonne Universit\'e and CNRS UMR 7589,
4 place Jussieu, 75252 Paris Cedex 05, France}

\author{Roderich Moessner}
\affiliation{Max Planck Institute for the Physics of Complex Systems, N\"othnitzer
Str. 38, 01187 Dresden, Germany}

\begin{abstract}
We present an analysis of the orbital Hall effect (OHE) in a strip geometry and derive a formula for the orbital angular momentum (OAM) accumulation at the edges. The result is expressed in terms of band structure parameters and scattering rates, providing a link between experimental observations of the OHE and the underlying microscopics. A key result is that the effective OAM decay rate follows a Dykonov-Perel-like scaling and is {\it inversely} proportional to the electron scattering rate, even if the latter is small.
Furthermore, investigating OAM transport in an inhomogeneous setting, we show that non-Ohmic flows and spatially varying electric fields result in contributions to the OHE which are distinct from the well known intrinsic and extrinsic mechanisms.
\end{abstract}

\maketitle

\section{Introduction}

Recent obserations \citep{choi2023_OHE_Titanium,lyalin2023_OHE_Chromium,sala2023orbital_Hanle} of the orbital Hall effect \citep{bernevig2006_OHE_original,kontani2009giant_OHE_transition,go2018intrinsic_orbital_textures} in different metals has stimulated a variety of research on the topic \citep{burgos2024_OHE_review,yang2024orbital_torque,busch2023_edge_OHE,tang_bauer_2024_disorder_intrinsic_OAM_hall,voss2025_edge_OHE_ballistic,go2024_first_principle_OHE,liu2024_culcer_extrinsic,liu2025quantum_corrections_OHE,fonseca2023_OHE_mesoscopic}. It has been demonstrated that the orbital Hall effect (OHE) can lead to large response \citep{jo2018gigantic_OHE_light_metals,choi2023_OHE_Titanium}, and that the dynamics of orbital magnetization can be distinct from its spin couterpart \citep{han2022orbital_dynamics,sohn2024_orbital_dyakonov_perel}. 

\begin{figure}
\vspace{0.4cm}
\centering
 \includegraphics[width=0.85\columnwidth]{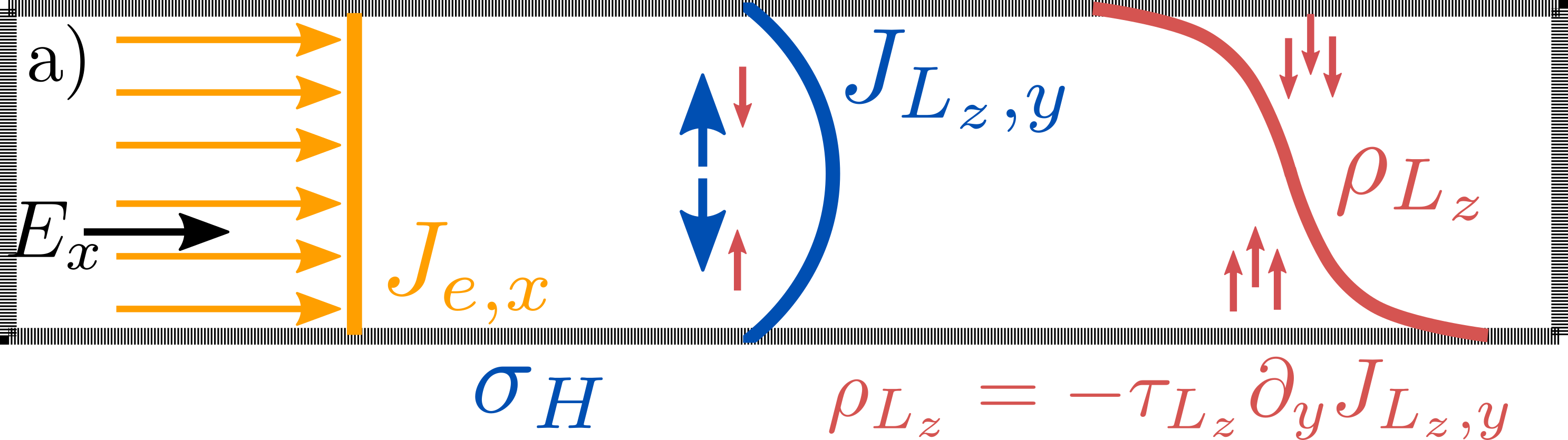}
 \caption{The Hall strip geometry studied in this paper. An electric field $E_x$ parallel to the $x$-axis creates a longitudinal charge current $J_{e,x}$ and a transverse orbital angular momentum current $J_{L_z,y}$. The latter induces excess orbital angular momentum (OAM) $\rho_{L_{z}}$ which is of opposite polarization at opposing edges of the strip. The accumulation of OAM in the steady state is governed by a "lossy" continuity equation: $\rho_{L_{z}}=-\tau_{L_{z}}\partial_{y}J_{L_{z},y}$, where $\tau_{L_{z}}$ is the effective decay time of OAM.}
\label{Hall_strips}
\end{figure}

Because of an intricate interplay of different time and length scales, scattering channels, non-trivial band geometry, and of course temperature, the microscopic modeling of orbital Hall effects is challenging. In experiment, geometric properties of the sample enter as additional parameters.  A detailed analysis of the OHE in a strip geometry (see Fig. \ref{Hall_strips}) that would relate key observables to microscopic parameters and transport mechanisms, such as intrinsic and extrinsic contributions, is still missing to our best knowledge.  

The obvious promise of a rich set of transport regimes, along with their current experimental relevance \citep{matsumoto2025observation_OHE_Si,gupta2025OHE_spin_orbit_torque,bhowal2021_OHE_graphene,fukunaga2023orbital,chen2024topology_OHE,veneri2025extrinsic_OHE_skew_transition}, clearly calls for an in-depth investigation of
orbital angular momentum transport in confined geometries, taking into account the effects of orbital angular momentum (OAM) diffusion and decay. Staring from a common model for the OHE, and treating scattering within a generalized relaxation time ansatz, we use the quantum Boltzmann approach to study the effects of confinement and OAM relaxation on the OHE. We focus on the weak scattering limit, where the inter-band distance is much larger then the inverse scattering time.

Within this framework, we discuss the main mechanisms of OAM relaxation (see Fig. \ref{relaxation_regimes}): the Elliott-Yafet mechanism, where the OAM decay rate is proportional to the electron scattering rate, and the Dyakonov-Perel mechanism, where the OAM decay rate is inversely proportional to the scattering rate. The latter is of particular importance insofar it has been demonstrated recently that the Dyakonov-Perel mechanism is relevant for OAM-transport even in centrosymmetric systems \citep{sohn2024_orbital_dyakonov_perel}.

Our main result is a formula for the accumulation of OAM on the edges of a Hall strip -- a central quantity in experiments on the OHE -- that links the OAM distribution to microscopic quantities such as scattering rates and parameters characterizing the band structure [see Eq. \eqref{eq:OAM_accumulation}]. On the way, we make a number of surprising observations: (i) the effective decay rate of OAM in the steady state obeys a Dyakonov-Perel scaling, even in the limit of weak scattering. This feature results from the interplay of the intrinsic orbital $k$-space texture and electron scattering. (ii) Parametrizing perturbations to the equilibrium density matrix in terms of angular harmonics of the polar angle of momentum $\varphi_\mathbf{k}$ [see Eqs. \eqref{eq:dens_mat_harm_expansion} and \eqref{eq:Collision_ansatz}], we find that it is crucial to distinguish between scattering rates for different angular harmonics. In particular, we show that, in the case of a single band crossing the Fermi energy, scattering on scalar impurities does not contribute to the decay of the accumulated density of OAM, and other scattering mechanisms, possibly otherwise sub-leading to impurity scattering, determine the effective relaxation rate of OAM. (iii) Extending the Quantum Boltzmann analysis to the spatially inhomogeneous case, i.e.\ more complex geometries, we observe several new mechanisms which can contribute to the OHE. We first observe that shear flows couple to OAM currents via what we call the non-local contribution to the orbital Hall conductivity (see Eqs \eqref{eq:g_(1)_bar_sol} and \eqref{eq:conductivity}). Secondly, we find that spatially non-uniform charge currents can lead to the accumulation of OAM [see Eq \eqref{eq:conti_rho_Lz}]. While these non-local effects do not contribute to the OHE in the case of uniform charge flow in an infinite Hall strip, we point to their possible importance under more complex conditions, e.g. for non-Ohmic/hydrodynamic flows as known to arise in ultra-clean materials \citep{gurzhi1963_hydro_1,gurzhi1968_hydro_2,deJong1995_GaAs_hydro,
moll2016_delafossite_hydro,levitov2016electron_vortices,
sulpizio2019_graphene_poiseuille,aharon2022_observation_vortices,
baker2024nonlocal_skin_effect,baker2024perspective_hydro,
wolf2023_parahydro}. We thus suggest that the interplay of non-Ohmic flows and OHE physics can lead to novel, geometrically-tunable effects.

\section{Modeling of orbital transport}

In this section we discuss our microscopic model for orbital transport.
Our results are, to a large extent, model independent. However, for
concreteness, we consider the tight binding Hamiltonian 

\begin{equation}
\hat{H}=
\left[\begin{array}{c}
c_{\alpha,\mathbf{k}}^{\dagger}\\
c_{\beta,\mathbf{k}}^{\dagger}
\end{array}\right]
\left[\begin{array}{cc}
\varepsilon_{\alpha}\left(\mathbf{k}\right) & H_{\alpha\beta}\left(\mathbf{k}\right)\\
H_{\alpha\beta}\left(\mathbf{k}\right) & \varepsilon_{\beta}\left(\mathbf{k}\right)
\end{array}\right]
\left[\begin{array}{c}
c_{\alpha,\mathbf{k}}\\
c_{\beta,\mathbf{k}}
\end{array}\right],
\label{eq:tight_binding_hamiltonian}
\end{equation}

where $c_{\alpha/\beta,\mathbf{k}}^{\dagger}$ and $c_{\alpha/\beta,\mathbf{k}}$
are creation and annihilation operators for two distinct orbitals
labeled by $\alpha$ and $\beta$. The Hamiltonian of Eq. (\ref{eq:tight_binding_hamiltonian})
provides a suitable minimal model for the OHE \citep{go2018intrinsic_orbital_textures,han2023_orbital_textures,tang_bauer_2024_disorder_intrinsic_OAM_hall}.

A variety of systems can be mapped on this Hamiltonian, in particular,
it can describe p-orbital systems if we choose $\alpha=p_{x}$ and
$\beta=p_{y}$, or transition metals if, instead, we associate $\alpha=d_{xz}$ and
$\beta=d_{yz}$ or $\alpha=d_{x^{2}-y^{2}}$ and $\beta=d_{xy}$ \citep{kontani2009giant_orbital_hall,han2023_orbital_textures,tang_bauer_2024_disorder_intrinsic_OAM_hall}.
The tight binding Hamiltonian (\ref{eq:tight_binding_hamiltonian})
can be represented in the classic form 
\begin{equation}
\hat{H}=\vec{c}_{\mathbf{k}}^{\dagger}\left[d_{0}\hat{\sigma}_{0}+\mathbf{d}\cdot\hat{\boldsymbol{\sigma}}\right]\vec{c}_{\mathbf{k}}
\label{}
\end{equation}
with $d_{0}=\left[\varepsilon_{\alpha}\left(k\right)+\varepsilon_{\beta}\left(k\right)\right]/2$,
$d_{1}=H_{\alpha\beta}$, $d_{2}=0$, and $d_{3}=\left[\varepsilon_{\alpha}\left(k\right)-\varepsilon_{\beta}\left(k\right)\right]/2$. Here and in the following the "hat" denotes matrices in orbital or band space.
We assume a time-reversal symmetric system and disregard spin orbit
coupling, such that all hopping parameters are real and therefore
$d_{2}=0$.

The eigenenergies of Eq.~(\ref{eq:tight_binding_hamiltonian}) are $\varepsilon_{\pm,\mathbf{k}}=d_{0}\left(\mathbf{k}\right)\pm\left|\mathbf{d}\left(\mathbf{k}\right)\right|$,
and the eigenstates are given by $\ket{+,\mathbf{k}}=\cos\left(\Theta_{\mathbf{k}}/2\right)\ket{\alpha}+\sin\left(\Theta_{\mathbf{k}}/2\right)\ket{\beta}$
and $\ket{-,\mathbf{k}}=\sin\left(\Theta_{\mathbf{k}}/2\right)\ket{\alpha}-\cos\left(\Theta_{\mathbf{k}}/2\right)\ket{\beta}$,
with $\Theta_{\mathbf{k}}=\arctan\left(d_{x}/d_{z}\right).$ In the following, we will refer to the basis $\ket{\pm,\mathbf{k}}$ as the band basis. Since, for the $p_x$, $p_y$ and $d_{xz}$, $d_{yz}$ pairs,
we have $\ket{L_{z}=\pm\hbar}=\mp\left(\ket{\alpha}\pm i\ket{\beta}\right)/\sqrt{2}$,
the OAM operator in the basis of Eq. (\ref{eq:tight_binding_hamiltonian}) is given by $\hat{L}_{z}=\hbar\hat{\sigma}_{y}$. In the band basis, we find
\begin{equation}
\hat{L}_{z}=-\hbar\hat{\sigma}_{y}.\label{eq:L_z}
\end{equation}
For the $d_{x^2-y^2}$, $d_{xy}$ pair,  the right hand side of Eq. \eqref{eq:L_z} is multiplied by an additional factor of two.

The intrinsic contribution to the OHE is sourced by the Berry connection
$\boldsymbol{\mathcal{A}}_{ab}=i\bra{a,\mathbf{k}}\frac{\partial}{\partial\mathbf{k}}\ket{b,\mathbf{k}}$,
where $a$, $b$ take the values  $\pm1$. In many models $\boldsymbol{\mathcal{A}}_{ab}$ only depends on $\varphi_{\mathbf{k}}$, the polar angle of $\mathbf{k}$,
and not its magnitude $k$. We will assume that $\frac{\partial}{\partial\varphi_{\mathbf{k}}}\Theta_{\mathbf{k}}=\nu$,
where $\nu$ is a constant, such that we find
\begin{equation}
\boldsymbol{\mathcal{A}}_{ab}=\frac{\nu i}{2k}\epsilon_{ab}\hat{\mathbf{e}}_{\varphi_{\mathbf{k}}}.\label{eq:berry_connection}
\end{equation}
This is true, e.g., for the widely used \citep{go2018intrinsic_orbital_textures,han2022orbital_dynamics,han2023_orbital_textures,tang_bauer_2024_disorder_intrinsic_OAM_hall}
model $d_{0}\sim k_{x}^{2}+k_{y}^{2}$, $d_{x}\sim2k_{y}k_{x}$ and
$d_{z}\sim k_{x}^{2}-k_{y}^{2}$ with $\nu=2$. Note that time-reversal symmetry constrains $d_1$ and $d_3$ to be even functions
of $\mathbf{k}$. Therefore, adding $\pi$ to $\varphi_{\mathbf{k}}$ changes $\Theta_{\mathbf{k}}$
by an integer multiple of $2\pi$. This forces $\nu$ to be an even integer.

To study transport of OAM, we employ the quantum Boltzmann equation.
In the eigenbasis, it reads
\begin{align}
 & \frac{\partial}{\partial t}\hat{f}+\frac{i}{\hbar}\left[\hat{h},\hat{f}\right]+\frac{1}{2}\left(\hat{\mathbf{v}}\cdot\nabla_{\mathbf{r}}\right)\hat{f}+\frac{1}{2}\left(\nabla_{\mathbf{r}}\hat{f}\right)\cdot\hat{\mathbf{v}}+\frac{e}{\hbar}\mathbf{E}\cdot\frac{\mathcal{D}\hat{f}}{\mathcal{D}\mathbf{k}}\nonumber \\
 & =\hat{C}\left[\hat{f}-\hat{f}_{\mathrm{eq}}\right].\label{eq:Quantum_Boltzmann}
\end{align}
Here, $\hat{C}$ is the collision operator, $\hat{h}_{ab}=\delta_{ab}\varepsilon_{b}\left(k\right)$
the diagonalized Hamiltonian, $\hat{f}_{ab}=\bra{a,\mathbf{k}}\hat{F}\left(\mathbf{k},\mathbf{r}\right)\ket{b,\mathbf{k}}$,
where $\hat{F}\left(\mathbf{k},\mathbf{r}\right)$ is the Wigner-tranformed
density matrix (see  \ref{subsec:Details-on-the}), and $\hat{\mathbf{v}}_{ab}=\left(\mathcal{D}\hat{h}/\mathcal{D}\mathbf{k}\right)_{ab}$
is the velocity operator, written in terms of the covariant derivative
$\mathcal{D}/\mathcal{D}\mathbf{k}$, defined as
\begin{equation}
\frac{\mathcal{D}\hat{X}}{\mathcal{D}\mathbf{k}}=\frac{\partial\hat{X}}{\partial\mathbf{k}}-i\left[\boldsymbol{\mathcal{A}},X\right]\label{eq:covariant_derivative}
\end{equation}
(see \ref{subsec:Details-on-the} and Ref. \citep{sekine2017quantum}). Explicitly, we find that the velocity operator is given by
\begin{equation}
    \mathbf{v}_{ab}=\hbar^{-1}\delta_{ab}\partial\varepsilon_{a}/\partial\mathbf{k}-i\hbar^{-1}\boldsymbol{\mathcal{A}}_{ab}\left(\varepsilon_{a}-\varepsilon_{b}\right)\,.
    \label{velocity}
\end{equation}

We divide the the density matrix $\hat{f}\left(\mathbf{k}\right)$
into an equilibrium and a non-equilibrium part: $\hat{f}\left(\mathbf{k}\right)=\hat{\mathbf{1}}f^{0}\left(k\right)+\delta\hat{f}\left(\mathbf{k}\right)$,
where $f^{0}\left(k\right)$ is the Fermi-Dirac distribution. For
what follows, it will be useful to expand the non-equilibrium part
in terms of angular harmonics of the polar angle of $\mathbf{k}$,
$\varphi_{\mathbf{k}}$:
\begin{equation}
    \delta\hat{f}\left(\mathbf{k}\right)=\sum_{m}\left(\hat{g}_{\left(m\right)}\cos m\varphi_{\mathbf{k}}+\hat{\bar{g}}_{\left(m\right)}\sin m\varphi_{\mathbf{k}}\right).\label{eq:dens_mat_harm_expansion}
\end{equation}

In what follows, we will assume that we are dealing with a strip that is infinite along the $x$-direction but has a finite width. The electric field is aligned with the $x$-axis: $\mathbf{E}=\hat{\mathbf{e}}_x$. Sample boundaries impose a spatial dependence on the transverse OAM
currents (see Fig.~\ref{Hall_strips}). Physical quantities of interest are the densities and currents of charge and OAM. Charge currents are found by tracing over the velocity operator:
\begin{equation}
    \mathbf{J}_{e}=\int\frac{d^2k}{(2\pi)^2}\mathrm{Tr}\left[e\mathbf{\hat{v}}\delta \hat{f}\right]\,, 
    \label{charge_current}
\end{equation}
while the transverse OAM currents can be found by tracing over the OAM current operator $\hat{\mathbf{j}}_{L_{z}}=\left\{ \mathbf{\hat{v}},\hat{L}_{z}\right\}$ \citep{liu2025_OAM_current_corrections}:
\begin{equation}
    \mathbf{J}_{L_z}=\int\frac{d^2k}{(2\pi)^2}\mathrm{Tr}\left[\left\{ \mathbf{\hat{v}},\hat{L}_{z}\right\}\delta \hat{f}\right]\,. 
    \label{orbital_current}
\end{equation}

Since $\hat{\mathbf{v}}$ is a vector in $k$-space, upon averaging over $\varphi_\mathbf{k}$, only the $m=1$ components  remain, which therefore encode all information about currents. Terms in $\delta\hat{f}$ that are proportional to $\cos{\varphi_\mathbf {k}}$,  contribute to currents that flow parallel to the electric field, while terms proportional to $\sin{\varphi_\mathbf {k}}$  source transverse currents.
OAM and charge densities, on the other hand, are determined by the $m=0$ components:
\begin{align}
    \rho_{e}=\int\frac{d^2k}{(2\pi)^2}\mathrm{Tr}\left[e\delta \hat{f}\right]\,,
    \label{charge_density}
    \\
    \rho_{L_z}=\int\frac{d^2k}{(2\pi)^2}\mathrm{Tr}\left[L_z\delta \hat{f}\right]\,.
    \label{orbital_density}
\end{align}
Since we will be dealing with inhomogeneous currents, and thus with
shear flows, we need to adequately describe shear induced momentum
transfer in our model. Therefore, in the spirit of a hydrodynamic
gradient expansion, we include harmonics up to $m=2$ \citep{huang2008statistical_mechanics}. 

\section{Scattering times}
We now discuss our treatment of electron scattering. We adopt a relaxation time approximation for the
collision operator $\hat{C}$. To keep the approach as general as possible, we do not specify the scattering mechanisms. Some properties of the collision operator are illustrated with the example of impurity scattering, but the relaxation times can also be associated with electron-phonon or electron-electron scattering. In a rotationally invariant system,
$\hat{C}$ does not mix channels with different $m$ in the expansion \eqref{eq:dens_mat_harm_expansion} (see Sec. \ref{subsec:scattering_times_symmetries}). However, $\hat{C}$
mixes the $\hat{g}_{\left(m\right)}$ and $\hat{\bar{g}}_{\left(m\right)}$
terms at any given $m$. In general, we can write 
\begin{widetext}
\begin{equation}
\hat{C}\left[\hat{g}_{\left(m\right)}\cos m\varphi_{\mathbf{k}}\right]_{ab} =-\sum_{cd}\left(\tau_{\left(m\right),ab|cd}^{-1}g_{\left(m\right),cd}\cos m\varphi_{\mathbf{k}}\right.
 \left.+\bar{\tau}_{\left(m\right),ab|cd}^{-1}g_{\left(m\right),cd}\sin m\varphi_{\mathbf{k}}\right)\ .
 \label{eq:Collision_ansatz}
\end{equation}
\end{widetext}
The action of the collision operator on $\hat{\bar{g}}_{\left(m\right)}\sin m\varphi_{\mathbf{k}}$ can be inferred by making the substitution $\varphi_{\mathbf{k}}\rightarrow\varphi_{\mathbf{k}}-\pi/2m$, for each $m$ (see \ref{subsec:scattering_times_symmetries}).

It should be noted that conservation laws impose constraints on the scattering rates. For example, momentum conservation, as in the case of electron-electron scattering, implies that the momentum integral over $\mathrm{Tr}\hbar\mathbf{k}\hat{C}_{\mathrm{el.}-\mathrm{el.}}[\delta\hat{f}]$ vanishes, which constrains the scattering rates for $m=1$. Similarly, charge conservation implies the vanishing of the momentum integral over $\mathrm{Tr}\hat{C}[\delta\hat{f}]$, which restricts scattering rates in the $m=0$ channel. In general, we assume that the three mechanisms of electron-impurity, electron-phonon, and electron-electron scattering are present, and the dominant mechanism has to be taken into account for each rate $\tau_{\left(m\right),ab|cd}^{-1}$. 

Some simplifications to the collision operator can be made based on symmetry considerations. 
Disregarding spatial derivatives and electric fields for the sake of this argument, we demand that $\delta \hat{f}$ be Hermitian at all times. By conjugating Eq. (\ref{eq:Quantum_Boltzmann}), and assuming real scattering rates \footnote{For scattering on scalar impurities, in the Born approximation, all scattering times are real. As shown in \ref{subsec:scattering_times_symmetries}, time-reversal symmetry and rotational invariance imply that most (except for example
$\tau_{\left(m\right),a\bar{a}|a\bar{a}}$ and $\bar{\tau}_{\left(m\right),a\bar{a}|\bar{a}a}$)
scattering rates that appear in the above collision operator are indeed real numbers.}, we find the relationships
\begin{align}
    \tau_{\left(m\right),ab|cd}&=\tau_{\left(m\right),ba|dc}
    \nonumber
    \\
    \bar{\tau}_{\left(m\right),ab|cd}&=\bar{\tau}_{\left(m\right),ba|dc}\,.
    \label{hermiticity_on_relax_times}
\end{align}

Furthermore, it is useful to assume that the system has a mirror symmetry in the $x$-$y$-plane. As demonstrated in \ref{subsec:scattering_times_symmetries} this enforces
\begin{equation}
    \bar{\tau}_{\left(m\right),aa|bb}^{-1}=\tau_{\left(m\right),a\bar{a}|bb}^{-1}=0\,,
    \label{symmetry_current_exclude}
\end{equation}
where $\bar{a}=-a$. Scattering rates of the type of Eq. \eqref{symmetry_current_exclude} would result in charge currents flowing orthogonal to the electric field, as well as to spin currents parallel to $\mathbf{E}$.  Both are prohibited by the mirror symmetry. 

Occasionally, when there is no ambiguity, we will use the abbreviation $\tau_{\left(m\right),aa|aa}^{-1}=\tau_{\left(m\right),a}^{-1}$ for the diagonal scattering times.

\section{Small parameters, spatial inhomogeneity and governing equations}

This section discusses small parameters stemming from the weak scattering assumption, as well as the effects of spatial inhomogeneity imposed by boundary conditions or an inhomogeneous electric field. We derive a closed set of equations which governs OAM transport in the strip geometry. Key simplifications are illustrated with the help of examples. For a full derivation we refer to \ref{subsec:full_set}.

\subsection{Scattering strength vs. band distance}

An important assumption throughout this paper is that the scattering rates are small compared to the band distance $\Delta\varepsilon_{\mathbf{k}}=\varepsilon_{+}(k)-\varepsilon_{-}(k)$. We parametrize this scale separation in terms of the small parameter 
\begin{equation}
\xi=\tau^{-1}\hbar/\Delta\varepsilon_{\mathbf{k}}\,,
\end{equation}
where $\tau$ stands for the relaxation time of any of the relevant
scattering channels (e.g. the smallest one). The presence of the small parameter $\xi$ will enable us to further simplify the scattering contributions. 

The reason for the appearance of $\Delta\varepsilon_{\mathbf{k}}$ is the commutator of the Hamiltonian with the density matrix in the second left hand side term of Eq. \eqref{eq:Quantum_Boltzmann}:
\begin{equation}
    \frac{i}\hbar{}\left[\hat{h},\hat{f}\right]_{ab} = a(1-\delta_{ab})\Delta\varepsilon_{\mathbf{k}}f_{ab}.
\end{equation}
Off-diagonal elements of the density matrix obtain a factor of $\pm\Delta\varepsilon_{\mathbf{k}}^{-1}$, while the diagonal elements are not affected. 

To analyze how the large band distance results in a scale separation of different contributions to the density matrix, it is useful to first disentangle  the action of the electric field on the components $\hat{g}_{(m)}$, $\hat{\bar{g}}_{(m)}$. Eqs. (\ref{eq:Quantum_Boltzmann}) and (\ref{eq:covariant_derivative}) indicate that the electric field couples to the momentum derivative of the density matrix $\hat{f}$ (first right hand side term of (\ref{eq:covariant_derivative})), as well as to the density matrix directly, through the Berry connection $\boldsymbol{\mathcal{A}}$. To linear order in $E$, the first contribution is
\begin{equation}
    \frac{e}{\hbar}\mathbf{E}\cdot\frac{\partial\varepsilon_{a,\mathbf{k}}}{\partial \mathbf{k}}\frac{\partial f_{0}\left(\varepsilon_{a,\mathbf{k}}\right)}{\partial\varepsilon_{a,\mathbf{k}}}\delta_{ab}
    \sim~\delta_{ab}\cos{\varphi_{\mathbf{k}}},
    \label{E-field_perturb_standard}
\end{equation}
while the second is given by
\begin{equation}
    i\frac{e}{\hbar}\mathbf{E}\cdot\boldsymbol{\mathcal{A}}_{ab}\left[f_{0}\left(\varepsilon_{a}\right)-f_{0}\left(\varepsilon_{b}\right)\right]\sim(1-\delta_{ab})\sin{\varphi_{\mathbf{k}}}.
    \label{Berry_E_source}
\end{equation}
When projecting the Quantum Boltzmann equation Eq. (\ref{eq:Quantum_Boltzmann}) onto the set of functions of Eq. (\ref{eq:dens_mat_harm_expansion}), a perturbation of the form of Eq. \eqref{E-field_perturb_standard}  couples to the diagonal components $g_{(1),aa}$ of $\delta \hat{f}$. Therefore, we find $g_{(1),aa}\sim \xi^0$. The off-diagonal components $\bar{g}_{(1),a\bar{a}}$ are sourced by both the Berry-connection contribution (Eq. \eqref{Berry_E_source}), and by scattering in the channels $\bar{\tau}^{-1}_{(1)a\bar{a}|bb}$, where the latter couple $\bar{g}_{(1),a\bar{a}}$ to $g_{(1),aa}$. Despite the factor of $\Delta\varepsilon_{\mathbf{k}}^{-1}$, $\bar{g}_{(1),a\bar{a}}$ is also of order $\xi^0$. This is because terms sourced by the Berry connection do not contribute a factor of $\tau^{-1}$, and $g_{(1),aa}\sim\tau$, so that the scattering rates cancel. In essence, this rephrases the well-known fact that the intrinsic (Berry-connection sourced) and extrinsic (scattering sourced) contributions to the orbital Hall effect are both of zeroth order in the scattering strength \citep{liu2024_culcer_extrinsic,tang_bauer_2024_disorder_intrinsic_OAM_hall}.  Simplifications arise when we deal with contributions of the type
\begin{align}
    \tau^{-1}_{(1)a\bar{a}|b\bar{b}}\bar{g}_{(1),b\bar{b}}&\sim\xi^1
    \\
    \bar{\tau}^{-1}_{(1)aa|b\bar{b}}\bar{g}_{(1),b\bar{b}}&\sim\xi^1\,,
\end{align}
as they result in sub-leading corrections to $\bar{g}_{(1),a\bar{a}}$ and $g_{(1),aa}$, respectively. Such terms are neglected below.

\subsection{Spatial inhomogeneity}

We now turn to the effects of inhomogeneity. In the presence of spatial variations, the Quantum Boltzmann  equation \eqref{eq:Quantum_Boltzmann} becomes non-diagonal in $m$. To see this, we perform a Fourier transform $\mathbf{r}\rightarrow\mathbf{q}$ of Eq. \eqref{eq:Quantum_Boltzmann}. Having the strip geometry in mind, we assume, as stated above, that the system is homogeneous along the $x$-axis and spatially varying in the transverse direction. This fixes the direction of the wavenumbers: $\mathbf{q}\propto\hat{\mathbf{e}}_y$. The spatial gradients in Eq. \eqref{eq:Quantum_Boltzmann}  then explicitly depend on $\varphi_\mathbf{k}$ via $\mathbf{q}\cdot\mathbf{v}_{aa}\sim\cos{\varphi_{\mathbf{k}}}$ and $\mathbf{q}\cdot\mathbf{v}_{a\bar{a}}\sim\sin{\varphi_{\mathbf{k}}}$ (see Eq. \eqref{velocity}). They thus couple angular harmonics with the index $m$ to their counterparts with indices $m+1$ and $m-1$. The characteristic time-scale for this coupling is given by 
\begin{equation}
    \tau^ {-1}_q=\max{||\hat{\mathbf{v}}(k)||}q\,,
\end{equation}
where $\max{||\hat{\mathbf{v}}(k)||}$ denotes the largest relevant velocity (because of the $\boldsymbol{\mathcal{A}}$-dependent contribution, velocities away from the Fermi-surface are also relevant). The relaxation time approximation is valid if $\tau^ {-1}_q\lesssim\tau^{-1}$, and we conclude that
\begin{equation}
    \tau^ {-1}_q\hbar/\Delta\varepsilon_\mathbf{k}\lesssim\xi ,
    \label{inhomogeneity_small}
\end{equation}
which leads to further simplifications. As an example consider $\hat{g}_{(1)}$ and $\hat{\bar{g}}_{(1)}$, which both couple to $\hat{\bar{g}}_{(2)}$ through the diagonal, and off-diagonal velocity, respectively. The contribution stemming from $\hat{\bar{g}}_{(1)}$ is suppressed by virtue of Eq. \eqref{inhomogeneity_small}, and will be ignored when we calculate $\hat{\bar{g}}_{(2)}$. Again, we refer to \ref{subsec:full_set} for a full analysis of the consequences of the smallness of $\tau^ {-1}_q\hbar/\Delta\varepsilon_\mathbf{k}$.

Finally, we  neglect
inter-band scattering between the diagonal density matrix elements, i.e. relaxation rates of the form
$\tau_{\left(m\right),aa|\bar{a}\bar{a}}^{-1}$. Scattering events
of this type can be incorporated at the cost of some additional algebra,
but do not lead to any interesting observations. In fact, as shown in \ref{subsec:inter-band-times}, for small $q$, their effect can be absorbed into a redefinition of the velocity components and scattering rates.

\subsection{Governing equations}

With the simplifications outlined above we are ready to write down a system of coupled equations for the components $\hat{g}_{(m)}$ and $\hat{\bar{g}}_{(m)}$ with $m<2$, which determine the steady-state of the system. We project the Quantum Boltzmann equation \eqref{eq:Quantum_Boltzmann} onto the set of basis functions of Eq. \eqref{eq:dens_mat_harm_expansion}, ignoring components with $m>2$ and terms of sub-leading order in $\xi$ and $\tau^ {-1}_q\hbar/\Delta\varepsilon_\mathbf{k}$. Exploiting the symmetry-imposed limitations of Eqs. \eqref{hermiticity_on_relax_times} and \eqref{symmetry_current_exclude}, in steady state ($\partial_{t}\hat{f}=0$), we obtain the following closed systems of coupled equations for the
matrix elements of $\hat{g}_{\left(m\right)}$, $\hat{\bar{g}}_{\left(m\right)}$
in the channels $m=0$, $m=1$ and $m=2$, respectively
\begin{widetext}
\begin{equation}
a\frac{i}{\hbar}\Delta\varepsilon_{\mathbf{k}}g_{\left(0\right),a\bar{a}}+\frac{iq}{4}\left(v_{aa}+v_{\bar{a}\bar{a}}\right)\bar{g}_{\left(1\right),a\bar{a}}+\frac{iq}{4}v_{a\bar{a}}\left(g_{\left(1\right),aa}+g_{\left(1\right),\bar{a}\bar{a}}\right)=-\sum_{b}\tau_{\left(0\right)a\bar{a}|b\bar{b}}^{-1}g_{\left(0\right),b\bar{b}},\label{eq:g_(0)_Eq_final}
\end{equation}
\begin{align}
\frac{iq}{2}v_{a}\bar{g}_{\left(2\right),aa}+\frac{e}{\hbar}E\frac{\partial\varepsilon_{a,\mathbf{k}}}{\partial k}\frac{\partial f_{0}\left(\varepsilon_{a,\mathbf{k}}\right)}{\partial\varepsilon_{a,\mathbf{k}}} & =-\tau_{\left(1\right),aa|aa}^{-1}g_{\left(1\right),aa},\label{eq:eq:g_(1)_Eq_final}
\end{align}
\begin{align}
a\frac{i}{\hbar}\Delta\varepsilon_{\mathbf{k}}\bar{g}_{\left(1\right),a\bar{a}}+\frac{iq}{4} v_{a\bar{a}} \left(\bar{g}_{\left(2\right),aa} + \bar{g}_{\left(2\right),\bar{a}\bar{a}}\right) + i\frac{e}{\hbar}E_{x}\mathcal{A}_{a\bar{a}}\left[f_{0}\left(\varepsilon_{a}\right)-f_{0}\left(\varepsilon_{\bar{a}}\right)\right] & =-\sum_{b}\bar{\tau}_{\left(1\right),a\bar{a}|bb}^{-1}g_{\left(1\right),bb},\label{eq:g_bar_(1)_Eq_final}
\end{align}
\begin{equation}
\frac{iq}{2}v_{a}g_{\left(1\right),aa}=-\tau_{\left(2\right),aa|aa}^{-1}\bar{g}_{\left(2\right),aa}.\label{eq:g_bar_(2)_Eq_final}
\end{equation}
\end{widetext}

Note that the ratio of the second and third terms in Eq. \eqref{eq:g_(0)_Eq_final} is of order $\xi$. We shall keep the second term, nevertheless. As we will see below, the third term is only non-zero for non-Ohmic charge flows with a curved current profile. For the typical case of an Ohmic flow, the second term is the leading contribution.

To keep the expression simple, we abbreviate $\tau_{\left(m\right),aa|aa}^{-1}=\tau_{\left(m\right),a}^{-1}$
and $v_{aa}=v_{a}$ in what follows. We note again that the symmetry arguments used here to simplify the scattering operator are discussed in more detail in \ref{subsec:scattering_times_symmetries}, while \ref{subsec:full_set} gives a more explicit derivation of Eqs. \eqref{eq:g_(0)_Eq_final} -- \eqref{eq:g_bar_(2)_Eq_final}.

\section{Intermezzo: Decay of Orbital Angular Momentum}
Having introduced the expansion of the quantum Boltzmann equation and scattering rates in terms of angular harmonics, we are now in the position to discuss the decay of the orbital angular momentum density, before coming back to its influence on the steady-state transport properties of the system. 

Angular momentum relaxation (both spin \citep{vzutic2004spintronics_review,fabian2007_semiconductor_spintronics_review,elliott1954_spin_relaxation,yafet1963_spin_relaxation,dyakonov_perel_1972_original1,dyakonov_perel_1984_original2} and orbital OAM \citep{sohn2024_orbital_dyakonov_perel}) typically obeys the Elliott-Yafet or Dyakonov-Perel scalings. In the Elliott-Yafet case, the angular momentum undergoes a damped precession, where the damping rate $\gamma$ is proportional to the scattering rate. In the Dyakonov-Perel case, the damping rate is inversely proportional to the scattering rate, such that the decay is slower for stronger scattering. For spin, Dyakonov-Perel type relaxation appears in non-centrosymmetric systems; however, it has been recently demonstrated, that for orbital systems, the Dyakonov-Perel scaling  is generic and appears even if central symmetry is preserved \citep{sohn2024_orbital_dyakonov_perel}. Our model captures  both Dyakonov-Perel and Elliott-Yafet mechanisms, which govern the decay of OAM in different regimes. 

Let us consider a spatially homogeneous system, in which the different channels $m$ are not coupled. For the two relevant scattering rates in Eq. \eqref{eq:Collision_ansatz}, we assume
$\tau_{\left(0\right)a\bar{a}|a\bar{a}}^{-1}=\tau_{\left(0\right)a\bar{a}|\bar{a}a}^{-1}=\tau_{\left(0\right)}^{-1}$. In \ref{subsec:Scattering-times} we show, that, for the case of impurity scattering, the two rates are indeed identical. Restoring the time-dependence, Eq. \eqref{eq:g_(0)_Eq_final} becomes
\begin{equation}
    \dot{g}_{\left(0\right),a\bar{a}}+a\frac{i}{\hbar}\Delta\varepsilon_{\mathbf{k}}g_{\left(0\right),a\bar{a}}=-\tau_{\left(0\right)}^{-1}g_{\left(0\right),a\bar{a}}-\tau_{\left(0\right)}^{-1}g_{\left(0\right),\bar{a}a}\,.
    \label{g_(0)_decay}
\end{equation}

\begin{figure}
\centering
 \includegraphics[width=1.0\columnwidth]{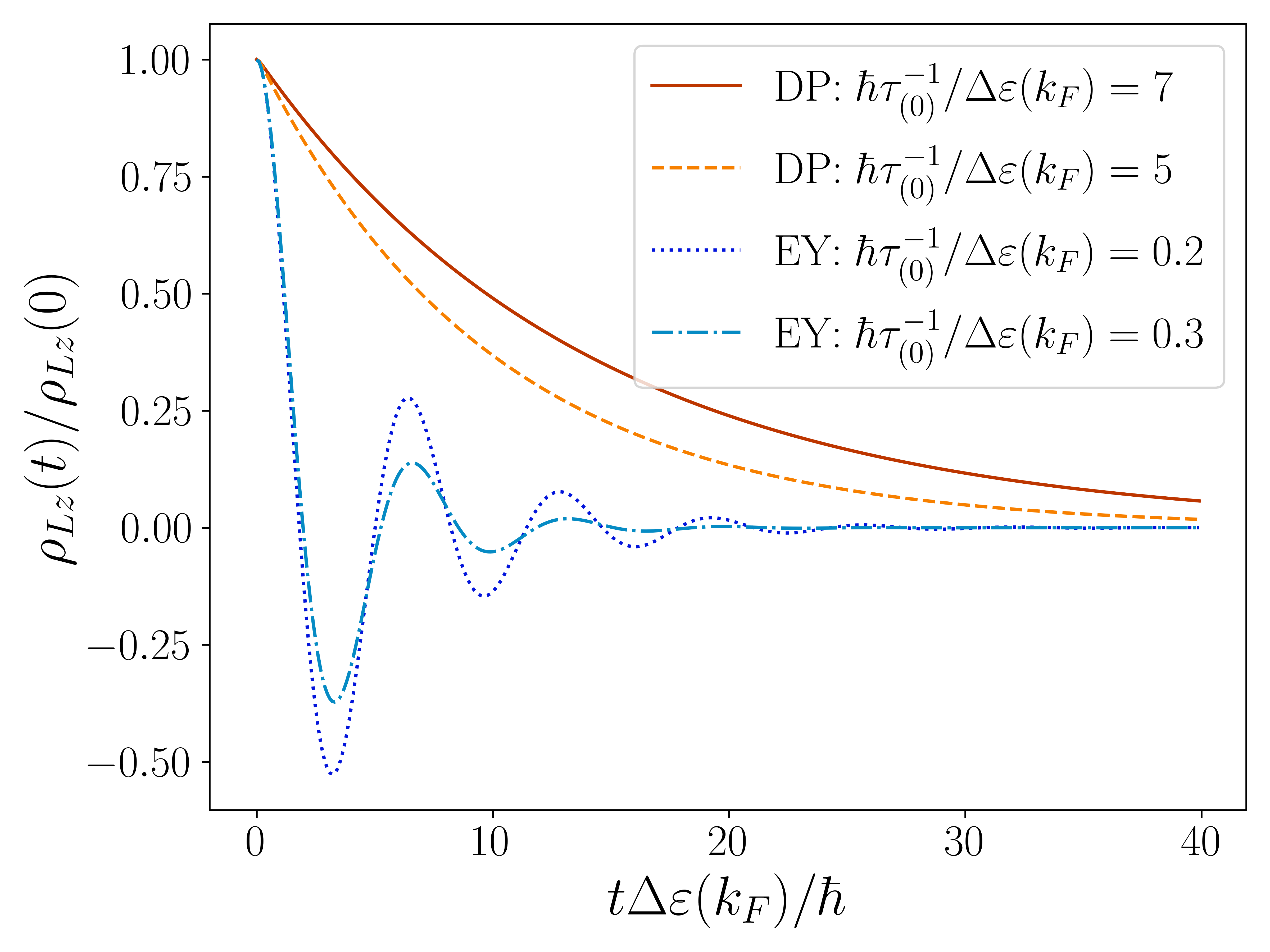}
 \caption{Relaxation of an initial orbital angular momentum (OAM) accumulation $\rho_{L_z}(0)$ in the Dyakonov-Perel (DP -- $\hbar\tau^{-1}_{(0)}\gg\Delta\varepsilon(k)$) and Elliott-Yaffet (EF -- $\hbar\tau^{-1}_{(0)}\ll\Delta\varepsilon(k)$) regimes according to Eq. \eqref{g_(0)_decay}. In the EF regime, the decay rate is proportional to the scattering rate $\tau^{-1}_{(0)}$,  whereas in the DP regime the rate is inversely proportional to the scattering rate.  Stronger scattering slows down the relaxation in the DP regime. In the steady state, the effective OAM decay rate follows  DP-scaling, irrespectively of the magnitude of $\hbar\tau^{-1}_{(0)}/\Delta\varepsilon(k)$ [see Eq. \eqref{eq:conti_rho_Lz}].}
\label{relaxation_regimes}
\end{figure}

Let us, for this section, also drop the assumption that $\Delta\varepsilon_\mathbf{k}\gg\hbar\tau^{-1}_{(0)}$.
Writing $g_{\left(0\right),a\bar{a}}=g_{\left(0\right),a\bar{a}}(0)e^{-\gamma t}$, we obtain the eigenvalues
\begin{equation}
\gamma=\tau_{\left(0\right)}^{-1}\pm\sqrt{\tau_{\left(0\right)}^{-2}-\Delta\varepsilon_{\mathbf{k}}^{2}/\hbar^{2}},\label{eq:density_decay_rate}
\end{equation}
The Elliott-Yafet damped, precessional
regime is found for $\hbar\tau_{\left(0\right)}^{-1}\ll\Delta\varepsilon_{\mathbf{k}}$
with $\gamma\approx\tau_{\left(0\right)}^{-1}\pm i\Delta\varepsilon_{\mathbf{k}}/\hbar$,
and the Dykonov-Perel (DP) regime for $\hbar\tau_{\left(0\right)}^{-1}\gg\Delta\varepsilon_{\mathbf{k}}$,
where, for long times, only the negative sign is relevant with $\gamma\approx\tau_{\left(0\right)}\Delta\varepsilon_{\mathbf{k}}^{2}/2\hbar^2$.
Notably, in the DP regime, the OAM relaxation rate grows when the
scattering rate $\tau_{\left(0\right)}^{-1}$ decreases -- a counterintuitive
behavior that originates from the interplay of precession and scattering. The different regimes of OAM relaxation are illustrated in Fig \ref{relaxation_regimes}.

For later purposes we define the relaxation time of OAM in the Dykonov-Perel regime
\begin{equation}
    \tau_{L_{z}}\left(k\right) =\frac{2\hbar^{2}\tau_{\left(0\right)}^{-1}}{\Delta\varepsilon^{2}_\mathbf{k}}\,.
\label{eq:effect_spin_relax}
\end{equation}
As is shown below, it is $\tau_{L_{z}}$ that takes the role of the effective OAM relaxation time in the steady state [see Eq. \eqref{eq:conti_rho_Lz}].  

We note that, for scattering off a scalar potential, in the
Born approximation, $\tau_{\left(0\right)}^{-1}=0$, if the Fermi energy
only cuts one of the two bands (see \ref{subsec:Scattering-times}). In the absence of inter-band
scattering, OAM relaxation through impurity scattering
is ineffective, and other scattering mechanisms 
dominate. For the transport behavior, this can have important consequences, because the scattering mechanisms controlling diffusion and decay can be distinct.

\section{Non-local orbital Hall conductivities}

The OAM response to an electric field is given by the orbital Hall conductivity
$\sigma_{H}$. For a strip of width $w$ extending infinitely in the  $x$-direction,
the current  depends on the $y$-coordinate. In Fourier space, we write 
\begin{equation}
J_{L_{z},y}\left(q\right)=\sigma_{H}\left(q\right)E_{x}(q)\label{eq:Ohm_mom_space}
\end{equation}
The orbital Hall current $J_{L_{z},y}\left(q\right)$ is found using Eq. \eqref{orbital_current}:
\begin{equation}
J_{L_{z},y}=\frac{i\hbar}{8\pi}\int\left(v_{++}+v_{--}\right)\left(\bar{g}_{\left(1\right)+-}-\bar{g}_{\left(1\right)-+}\right)kdk.\label{eq:OAM_current_explicit}
\end{equation}
Let us now solve for $\bar{g}_{\left(1\right),a\bar{a}}$. In the
first step, we combine Eqs. (\ref{eq:eq:g_(1)_Eq_final}) and (\ref{eq:g_bar_(2)_Eq_final})
to find 
\begin{equation}
g_{\left(1\right),aa}=-\frac{\tau_{\left(1\right),a}eEv_{a}}{1+\frac{1}{4}v_{a}^{2}q^{2}\tau_{\left(1\right),a}\tau_{\left(2\right),a}}\frac{\partial f_{0}\left(\varepsilon_{a}\right)}{\partial\varepsilon_{a}}.
\label{g_diag}
\end{equation}
Inserting this result back into Eq. (\ref{eq:eq:g_(1)_Eq_final}),
we find an expression for $g_{\left(1\right),aa}$, which can be used
to invert Eq. (\ref{eq:g_bar_(1)_Eq_final}). All in all, we find
that $\hat{\bar{g}}_{\left(1\right),a\bar{a}}$ is comprised of three
contributions: the intrinsic, the extrinsic, and what we call the
non-local contribution:
\begin{equation}
\hat{\bar{g}}_{\left(1\right),a\bar{a}}=\hat{\bar{g}}_{\left(1\right),a\bar{a}}^{\mathrm{int.}}+\hat{\bar{g}}_{\left(1\right),a\bar{a}}^{\mathrm{ext.}}+\hat{\bar{g}}_{\left(1\right),a\bar{a}}^{\mathrm{n.l.}}\label{eq:g_(1)_bar_sol}.
\end{equation}
Explicitly, the contributions are given by
\begin{align}
\hat{\bar{g}}_{\left(1\right),a\bar{a}}^{\mathrm{int.}} & =-\frac{a\hbar}{\left(\varepsilon_{+}-\varepsilon_{-}\right)}\frac{e}{\hbar}E_{x}\mathcal{A}_{a\bar{a}}\left[f_{0}\left(\varepsilon_{a}\right)-f_{0}\left(\varepsilon_{\bar{a}}\right)\right]\\
\hat{\bar{g}}_{\left(1\right),a\bar{a}}^{\mathrm{ext.}} & =\frac{ia\hbar}{\left(\varepsilon_{+}-\varepsilon_{-}\right)}\sum_{b}\bar{\tau}_{\left(1\right),a\bar{a}|bb}^{-1}g_{\left(1\right),bb}\\
\hat{\bar{g}}_{\left(1\right),a\bar{a}}^{\mathrm{n.l.}} & =-\frac{a\hbar}{\left(\varepsilon_{+}-\varepsilon_{-}\right)}\frac{q}{4}v_{a\bar{a}}\left(g_{\left(2\right),aa}+g_{\left(2\right),\bar{a}\bar{a}}\right).
\end{align}
It is interesting to note that the non-local contribution is sourced
by a perturbation in the $m=2$ channel, which, in the hydrodynamic context, describes the effects
of viscous shear. We will see below that in the simple strip geometry with Ohmic charge currents, this contribution does not effect the OAM current. It can, however, play a role in more complex geometries. We also note that in inverting for $g_{\left(1\right),aa}$ in Eq. \eqref{g_diag} , we ignore homogeneous solutions of Eqs. \eqref{eq:eq:g_(1)_Eq_final} and \eqref{eq:g_bar_(2)_Eq_final}, which hold for $E_x=0$. These are irrelevant for Ohmic flows, but are important for including non-trivial boundary conditions for curved flow profiles \citep{kiselev2020nonlocal,ledwith2019tomographic}. 

From Eqs. (\ref{eq:OAM_current_explicit}) and (\ref{eq:g_(1)_bar_sol})
we deduce that the orbital Hall conductivity is given by
\begin{align}
\sigma_{H}\left(q\right) & =\sigma_{H,0}^{\mathrm{int.}}+\sum_{a}\left(\frac{\sigma_{H,0}^{\mathrm{ext.\left(a\right)}}}{1+\frac{1}{4}v_{a,F}^{2}q^{2}\tau_{\left(1\right),a}\tau_{\left(2\right),a}}\right.\nonumber \\
 & \left.+\sigma_{H,0}^{\mathrm{n.l.\left(a\right)}}\frac{v_{a,F}^{2}q^{2}\tau_{\left(1\right),a}\tau_{\left(2\right),a}}{1+\frac{1}{4}v_{a,F}^{2}q^{2}\tau_{\left(1\right),a}\tau_{\left(2\right),a}}\right),\label{eq:conductivity}
\end{align}
where the index $a$ marks the contributions of the two bands. The
Hall conductivity consists of the intrinsic, the extrinsic and the
non-local contributions. In general, the expressions for $\sigma_{H,0}^{\mathrm{int.}}$,
$\sigma_{H,0}^{\mathrm{ext.}}$ and $\sigma_{H,0}^{\mathrm{n.l.}}$
depend on details of the chosen model. From Eq. (\ref{eq:OAM_current_explicit}),
we find, at low temperatures, $\sigma_{H,0}^{\mathrm{int.}}=\frac{e\nu\hbar}{4\pi}\int\frac{v_{+}+v_{-}}{\varepsilon_{+}-\varepsilon_{-}}E_{x}\left[f_{0}\left(\varepsilon_{+}\right)-f_{0}\left(\varepsilon_{-}\right)\right]dk$,
$\sigma_{H,0}^{\mathrm{ext.\left(a\right)}}=\frac{1}{4\pi}\frac{\tau_{\left(1\right),a}}{\bar{\tau}_{\left(1\right),+-|aa}}\frac{e\hbar k_{F,a}\left[v_{a,F}+v_{\bar{a}}\left(k_{F,a}\right)\right]}{\varepsilon_{F}-\varepsilon_{-}\left(k_{F,a}\right)}$,
as well as $\sigma_{H,0}^{\mathrm{n.l.\left(a\right)}}=\frac{e\nu}{16\pi}\left(1+\frac{v_{\bar{a}}\left(k_{F,a}\right)}{v_{a,F}}\right)$.
Here $\hbar k_{F,a}$ is the Fermi momentum of band $a$. In the simplest
case, the Fermi energy cuts only one of the bands, and the sum over
bands can be omitted. It is noteworthy that the extrinsic and non-local contributions stem from electrons at the Fermi surfaces, whereas the intrinsic contribution -- stemming from virtual hopping between the bands -- involves an extensive integral over regions where the lower band is occupied, while the upper band is empty.

Interestingly, the extrinsic contribution vanishes in the limit $q\neq 0$, $\tau_{(m),a}\rightarrow \infty$, whereas the non-local contribution approaches a fixed value. We will see below that this behavior could be consequential in strongly inhomogeneous electric fields.

\section{Diffusion of orbital angular momentum}

We now present an analysis of orbital Hall currents and the diffusion of OAM in a spatially inhomogeneous system and derive the $q$-dependent Hall conductivities. The steady state distribution of OAM currents in a strip geometry
including diffusive effects can be found using Eq. (\ref{eq:conductivity}). For simplicity, we assume that the Fermi
energy lies in the upper band, and the lower band is fully occupied.
We then confine ourselves to $a=+1$ in Eq. (\ref{eq:conductivity})
and drop the index $a$. The generalization to the case where the
Fermi energy cuts both bands is trivial, if inter-band scattering effects on the charge conductivity are neglected \footnote{These effects do not lead to qualitative changes in the orbital Hall conductivity, but can be absorbed in a redefinition of the Fermi velocities and scattering times (see \ref{subsec:inter-band-times})}.  Introducing the mean scattering
length 
\begin{equation}
    \bar{l}=\frac{1}{2}v_{F}\sqrt{\tau_{\left(1\right),+}\tau_{\left(2\right),+}}\,,
\end{equation}
we investigate the limits (i) $w/\bar{l}\gg1$, where the sample width
is large compared to the scattering length, and (ii) $w/\bar{l}\ll1$
where the sample width is smaller than the scattering length. We remind the reader that $\tau_{(m),+}$ is a shorthand for the scattering time $\tau_{(m),++|++}$. While case (ii) somewhat overstretches the relaxation time approximation, it serves as a useful illustration of the behavior in ultra-clean samples.

Since we are interested in the OAM current distribution given an electric
field $E(y)$, we have to invert Eq. (\ref{eq:Ohm_mom_space}). This can be achieved by multiplying by the common denominator in Eq. \eqref{eq:conductivity}. Doing so, and subsequently performing an inverse Fourier transform we find the equation

\begin{equation}
J_{L_{z}}''\left(y\right)-\frac{1}{\bar{l}^{2}}J_{L_{z}}(y)=-\frac{\sigma_{1}}{\bar{l}^{2}}E_{x}(y)+\sigma_{2}E_{x}''(y),
\label{eq:current_diffusion}
\end{equation}
where we have abbreviated
\begin{align}
\nonumber
\sigma_{1} & =\sigma_{H,0}^{\mathrm{ext.}}+\sigma_{H,0}^{\mathrm{int.}}\\
\sigma_{2} & =\sigma_{H,0}^{\mathrm{int.}}+4\sigma_{H,0}^{\mathrm{n.l.}}\,.
\end{align}
In \ref{subsec:Derivation-of-the_diff}, we present a more technical derivation of Eq. \eqref{eq:current_diffusion} taking into consideration the finite width of the strip.

It must be noted that a static field $E_x(y)$ that is uniform in the $x$-direction and varying in the $y$-direction cannot be written as the gradient of a scalar potential and therefore, strictly speaking, is unphysical. However, in more complex geometries strongly inhomogeneous fields that are spatially varying in both directions are realistic. We therefore keep the second right hand side term in Eq. \eqref{eq:current_diffusion} for demonstrational purposes. While we will consider a uniform field in what follows, we note that, because of the dominant behavior of the nonlocal contribution to the orbital Hall conductivity $\sigma_{H,0}^{\mathrm{n.l.}}$ in the limit $q\rightarrow\infty$ (see Eq. \eqref{eq:conductivity}), its effects could be relevant in more complex settings. 

We now focus on Ohmic charge currents where $E_{x}=\mathrm{const.}$
Solving Eq. (\ref{eq:current_diffusion}) with the boundary conditions
$J_{L_{z}}\left(\pm w/2\right)=0$, we find
\begin{equation}
J_{L_{z}}\left(y\right)=\sigma_{1}E_{x}\left(1-\frac{\cosh\left(y/\bar{l}\right)}{\cosh\left(w/2\bar{l}\right)}\right).\label{eq:current_solution}
\end{equation}
For $w/\bar{l}\gg1$, the orbital Hall current density approaches
a nearly uniform flow with $J_{L_{z}}\left(y\right)=\sigma_{1}E_{x}$ within
the sample boundaries. Diffusion and inhomogeneity play a subordinate
role. For $w/\bar{l}\ll 1$, on the other hand, we find
\begin{equation}
J_{L_{z}}\left(y\right)=\frac{\sigma_{1}E_{x}}{\bar{l}^{2}}\left(\frac{w^{2}}{8}-\frac{y^{2}}{2}\right).\label{eq:current_l>>w}
\end{equation}
This distribution of $J_{L_{z}}\left(y\right)$ is strongly non-uniform.
For $\bar{l}^{2}\rightarrow\infty$, Eq. (\ref{eq:current_l>>w})
indicates the inefficient generation of orbital Hall currents in a
confined geometry. This is a consequence of the fact that inducing
a finite curvature $J_{L_{z}}''\left(y\right)$ by a uniform field
in Eq. (\ref{eq:current_diffusion}) becomes increasingly difficult
at large $\bar{l}$. This suppression is avoided for a non-uniform
electric field, since the contributions $J_{L_{z}}''\left(y\right)$
and $\sigma_{2}E_{x}''\left(y\right)$ are both of order $\mathcal{O}\left(\bar{l}^{0}\right)$. Interestingly, $\sigma_2$ does not involve the extrinsic contribution. However, to explore these qualitative differences for the OHE in uniform and non-uniform fields, more complex geometries must be studied. We leave this topic for future investigations and turn to the accumulation of OAM in the Hall strip.

\section{Accumulation of orbital angular momentum}

In this section we connect our previous analysis of orbital Hall currents to the accumulation of OAM across the sample -- a crucial quantity in the experimental observation of the OHE. We calculate the steady-state
OAM density according to Eq. \eqref{orbital_density}
from the Quantum Boltzmann equation. Using Eq. (\ref{eq:g_(0)_Eq_final}),
we can express $\rho_{L_{z}}\left(y\right)$ in terms of the OAM current
$J_{L_{z}}\left(y\right)$ and the electric current $J_{e,x}$$\left(y\right)$.
It is useful to introduce the auxiliary quantity $J_{L_{z},k,y}=\mathrm{Tr}\left[\hat{\bar{g}}_{\left(1\right)}\left\{ \hat{v}_{y},\hat{L}_{z}\right\} \right]$, such that the full OAM current is obtained by taking an integral
over $k$. Similarly, we define the OAM density at given $k$, $\rho_{L_z,k}$, and the charge current $J_{e,k,x}$. It should be kept in mind that these quantities also depend
on $y$. For simplicity, we again assume that the Fermi energy lies
in the upper band only. Inverting Eq. (\ref{eq:g_(0)_Eq_final}) to
find $\hat{g}_{\left(0\right)}$, we obtain the expression
\begin{equation}
\rho_{L_{z},k}=-iq\tau_{L_{z}}\left(k\right)J_{L_{z},k,y}+iq\tau_{L_{z},e}\left(k\right)\frac{\hbar}{e}J_{e,k,x},\label{eq:k_space_continuity_before_fourier}
\end{equation}
which relates the OAM density to the OAM and electrical currents at
each $k$. Here, $\tau_{L_{z}}$ is given by the Dyakonov-Perel relaxation time of Eq. \eqref{eq:effect_spin_relax}, and we have introduced.
\begin{equation}
    \tau_{L_{z},e}\left(k\right) =\frac{\hbar}{\Delta\varepsilon_\mathbf{k}}\frac{v_{+-}\left(k\right)}{v_{+}\left(k\right)}.
\label{eq:effect_spin_relax_current}
\end{equation}

Integrating formula (\ref{eq:k_space_continuity_before_fourier})
over $k$ is unproblematic for the extrinsic and nonlocal contributions,
since the integrand is localized at $k=k_{F}$. The intrinsic contribution,
however, contains an extended integral over $k$. Simplifying, we
assume that the band distance is approximately constant for $k>k_{F}$,
i.e. $\Delta\varepsilon\left(k\right)\approx\Delta\varepsilon\left(k_{F}\right)$.
This allows us to write the relationship between the total $\rho_{L_{z}}$,
$J_{L_{z},y}$, and $J_{e,x}$ in a closed form:
\begin{equation}
    \rho_{L_{z}}=-\tau_{L_{z}}\partial_{y}J_{L_{z},y}+\tau_{L_{z},e}\partial_{y}\frac{\hbar}{e}J_{e,x},\label{eq:conti_rho_Lz}
\end{equation}
where $\tau_{L_{z}}=\tau_{L_{z}}\left(k_{F}\right)$ and $\tau_{L_{z},e}=\tau_{L_{z},e}\left(k_{F}\right)$. The second right hand side term is relevant for inhomogeneous electric current flows, e.g. if the electric conduction is non-Ohmic. This term is induced by off-diagonal components of the velocity operator (see Eq. \eqref{eq:effect_spin_relax_current}), and offers an alternative mechanism for accumulating OAM. We leave the exploration of this term, as we did with the non-local contributions to the OAM Hall current, for future studies, and focus on
Ohmic behavior, i.e. a uniform electric current profile.
The last term in the above equation then vanishes.

The remainder of the formula
can be interpreted as a continuity equation with a decay time $\tau_{L_{z}}$
for the OAM density: $\partial_{t}\rho_{L_{z}}=-\tau^{-1}_{L_{z}}\rho_{L_{z}}-\partial_{y}J_{L_{z},k,y}$, assuming $\partial_{t}\rho_{L_{z}}=0$ (steady state).
Together with Eq. (\ref{eq:current_solution}), Eq. (\ref{eq:conti_rho_Lz})
can be used to calculate the OAM density profile:

\begin{equation}
\rho_{L_{z}}\left(y\right)=\tau_{L_{z}}\sigma_{1}E_{x}\frac{\sinh\left(y/\bar{l}\right)}{\cosh\left(w/2\bar{l}\right)}.
\label{eq:OAM_accumulation}
\end{equation}

\begin{figure}
\centering
 \includegraphics[width=1\columnwidth]{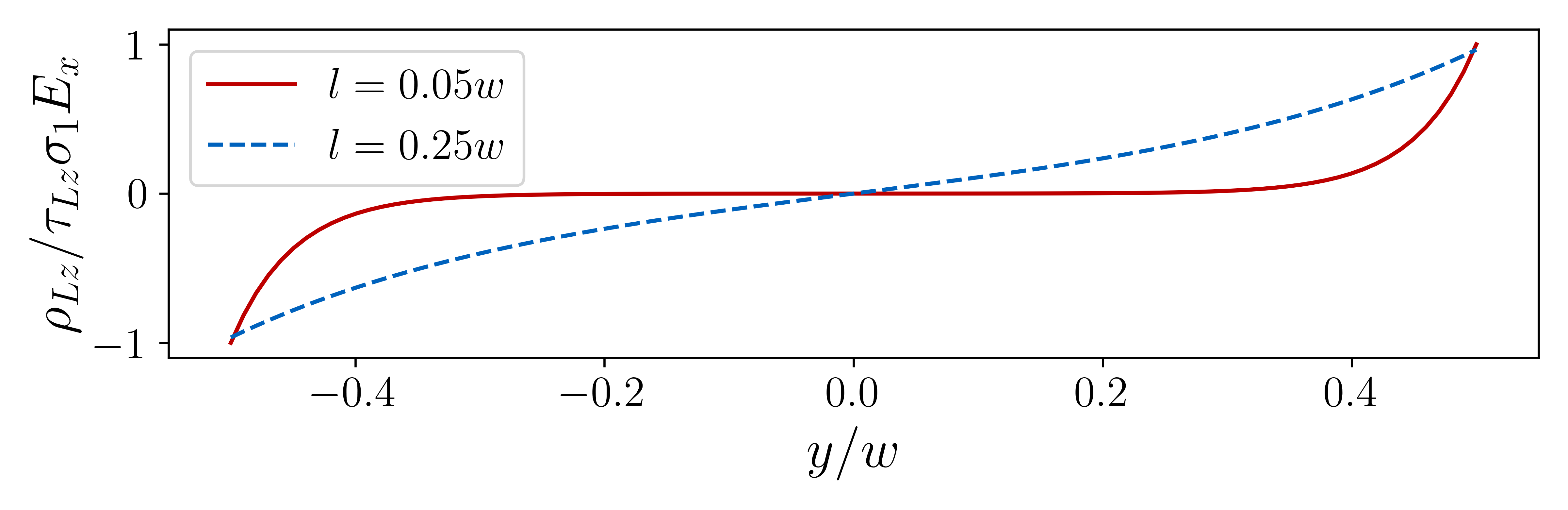}
\caption{Profiles of accumulated orbital angular momentum in a Hall strip (see Eq. \eqref{eq:OAM_accumulation}). The ratio $\bar{l}/w$ shapes the curvature of the profile.}
\label{profile}
\end{figure}

Some comments are due with regard to formula \eqref{eq:OAM_accumulation} and related phenomenological results from spin Hall physics \citep{zhang2000spin_Hall_diffusion}. Eq. \eqref{eq:OAM_accumulation} links the OAM accumulation to the microscopic parameters $\tau_{\left(m\right)}$
and $\Delta\varepsilon(k_F)$. In particular, it includes Dyakonov-Perel
physics. In the steady state, the interplay between OAM precession and decay results in the effective OAM
decay time of Eq. (\ref{eq:effect_spin_relax}), where $\tau_{L_{z}}\left(k\right)\sim\tau_{\left(0\right)}^{-1}$.

This is in contrast to approaches where no distinction between the channels $m$ is made, which can lead to grave consequences. For example, as pointed out above, in systems with a single band crossing the Fermi energy, impurity scattering does not induce a finite $\tau_{\left(0\right)}^{-1}$ and other mechanisms (e.g. electron-phonon or electron-electron scattering), which can be sub-leading in channels with $m>0$, can determine the magnitude of the OAM accumulation, as well as other properties, such as its temperature dependence. Another observation is that,
in the steady state, the fact that the effective decay time is proportional
to the impurity scattering rate in the $m=0$ channel is valid
for $\hbar\tau_{\left(0\right)}^{-1}\ll\Delta\varepsilon_{\mathbf{k}}$ and
 Dyakonov-Perel scaling therefore holds in the weak scattering regime.

\section{Discussion}

We have analyzed the orbital Hall effect in a confined geometry and
in inhomogeneous fields, taking into account the diffusion of orbital
angular momentum and non-local effects. We have derived formulas for the spatial profiles of
the orbital angular momentum density and the orbital Hall current
in a strip geometry, linking them to microscopic parameters. This revealed unusual relaxational properties of the OAM density, such as the Dyakonov-Perel scaling of the effective relaxation time. In addition,
we pointed to a surprising effect that can appear for non-Ohmic electrical currents: gradients of the electric current density contribute to the accumulated orbital momentum -- a phenomenon that is induced by the off-diagonal elements of the velocity operator. Non-Ohmic flows appear in ultra-clean materials \citep{gurzhi1963_hydro_1,gurzhi1968_hydro_2,deJong1995_GaAs_hydro,
moll2016_delafossite_hydro,levitov2016electron_vortices,
sulpizio2019_graphene_poiseuille,aharon2022_observation_vortices,
baker2024nonlocal_skin_effect,baker2024perspective_hydro,
wolf2023_parahydro}, and are strongly dependent on boundary conditions \citep{kiselev2019boundary,kiselev2019boundary,moessner2019boundary_engineering_piotr,
wolf2023_parahydro}. This hints at the possibility of unconventional orbital Hall response in tailored sample geometries, with nontrivial interplay between the vorticity of electrical currents \citep{aharon2022_observation_vortices,nazaryan2024nonlocal_vortices,
doornenbal2019spin_orbit_hydro_vortices} and Hall magnetization.

\begin{acknowledgments}
We acknowledge useful discussions with Graham Baker and Andrew Mackenzie. This project has received
funding from the European Union's Horizon 2023 research
and innovation programme under the Marie Sk\l odowska-Curie Grant Agreement No. 101155351 and the Deutsche Forschungsgemeinschaft under cluster of excellence ct.qmat (EXC 2147, project-id 390858490).
\end{acknowledgments}

\newpage
\onecolumngrid
\section*{Appendix} 

\renewcommand{\thesection}{Appendix}%

\setcounter{figure}{0}
\renewcommand{\thefigure}{C\ \arabic{figure}}%

\setcounter{equation}{0}
\renewcommand{\theequation}{A\,\arabic{equation}}%

\subsection{Details on the derivation of the quantum Boltzmann equation\label{subsec:Details-on-the}}

In this section we rationalize the appearance of convariant derivatives
and the Berry connection in the quantum Boltzmann equation (\ref{eq:Quantum_Boltzmann}).
Let $\hat{\rho}_{\mathbf{k}}$ be the density matrix in the orbital
basis. The Quantum Boltzmann equation is derived from the von-Neumann
equation
\begin{equation}
i\hbar\frac{\partial}{\partial t}\hat{\rho}_{\mathbf{k},\mathbf{k}'}=\left[H\left(\mathbf{k}\right)+eV\left(\mathbf{k}\right),\hat{\rho}_{\mathbf{k},\mathbf{k}'}\right],\label{eq:von_Neumann}
\end{equation}
where $V_{\mathbf{k},\mathbf{k}'}$ stemms from the appliead electric
fields. Then the Wigner transform of $\hat{\rho}_{\mathbf{k}}$, $\hat{F}\left(\mathbf{k},\mathbf{r}\right)$,
appearing below Eq. (\ref{eq:Quantum_Boltzmann}), is given by 
\begin{equation}
\hat{F}\left(\mathbf{k},\mathbf{r}\right)=\int\frac{d^{d}q}{\left(2\pi\right)^{d}}e^{i\mathbf{q}\cdot\mathbf{r}}\mathrm{\hat{\rho}_{\mathbf{k}+\mathbf{q}/2,\mathbf{k}-\mathbf{q}/2}}.
\end{equation}
The left hand side of Eq. (\ref{eq:Quantum_Boltzmann}) is then derived
by writing Eq. (\ref{eq:von_Neumann}) in terms of $\hat{F}\left(\mathbf{k},\mathbf{r}\right)$,
expanding in $\mathbf{q}$, and going to the eigenbasis of the Hamlitonian
(\ref{eq:tight_binding_hamiltonian}). Some care is needed for terms
that involve a derivative with respect to $\mathbf{k}$. The expression
$\mathcal{D}\hat{X}/\mathcal{D}\mathbf{k}$ (see Eq. (\ref{eq:covariant_derivative}))
indicates a covariant derivative that takes into account the $\mathbf{k}$-dependence
of the basis vectors, when taking the $\mathbf{k}$ derivative. Let
$X_{ab}=\bra{a,\mathbf{k}}X\left(\mathbf{k},\mathbf{r}\right)\ket{b,\mathbf{k}}$
be matrix elements of $X$ in the basis of eigenstates $\ket{a,\mathbf{k}}$
and $\boldsymbol{\mathcal{A}}_{ab}=i\bra{a,\mathbf{k}}\frac{\partial}{\partial\mathbf{k}}\ket{b,\mathbf{k}}$
be the Berry connection, then
\begin{align}
\bra{a,\mathbf{k}}\frac{\partial\hat{X}}{\partial\mathbf{k}}\ket{b,\mathbf{k}} & =\bra{a,\mathbf{k}}\frac{\partial}{\partial\mathbf{k}}\left(\sum_{op}X_{op}\left(\mathbf{k}\right)\ket{o,\mathbf{k}}\bra{p,\mathbf{k}}\right)\ket{b,\mathbf{k}}\nonumber \\
 & \frac{\partial X_{ab}}{\partial\mathbf{k}}+\bra{a,\mathbf{k}}\sum_{op}X_{op}\left(\frac{\partial\ket{o,\mathbf{k}}}{\partial\mathbf{k}}\bra{p,\mathbf{k}}+\ket{o,\mathbf{k}}\frac{\partial\bra{p,\mathbf{k}}}{\partial\mathbf{k}}\right)\ket{b,\mathbf{k}}\nonumber \\
 & =\frac{\partial X_{ab}}{\partial\mathbf{k}}-i\sum_{op}X_{op}\left(\boldsymbol{\mathcal{A}}_{ao}\delta_{pb}-\delta_{ao}\boldsymbol{\mathcal{A}}_{pb}\right)\nonumber \\
 & =\frac{\partial X_{ab}}{\partial\mathbf{k}}-i\sum_{o}\boldsymbol{\mathcal{A}}_{ao}X_{ob}+i\sum_{p}X_{ap}\boldsymbol{\mathcal{A}}_{pb}\nonumber \\
 & =\frac{\partial X_{ab}}{\partial\mathbf{k}}-i\left[\boldsymbol{\mathcal{A}},X\right]_{ab}.\label{eq:cov_der_derive}
\end{align}
The collision operator for a two-band system such as the one described
by Eq. (\ref{eq:tight_binding_hamiltonian}) has been derived in Refs.
\citep{khaetskii2006nonexistence,culcer2017interband,tang_bauer_2024_disorder_intrinsic_OAM_hall}.

\subsection{Relaxation rates for impurity scattering\label{subsec:Scattering-times}}

In this section we substantiate our assumptions on the disorder scattering
times. For the Hamiltonian of Eq. (\ref{eq:tight_binding_hamiltonian}),
the collision operator is given by \citep{tang_bauer_2024_disorder_intrinsic_OAM_hall,khaetskii2006nonexistence,culcer2017_coherence_transport} :
\begin{align}
I_{\mathrm{coll}}\left[\hat{\rho}\right]\left(\mathbf{k}\right)_{ab} & =\frac{\pi}{\hbar}\int\frac{d^{2}k'}{\left(2\pi\right)^{2}}\sum_{cd}\left\{ \left[\delta\left(\varepsilon_{c}\left(\mathbf{k}'\right)-\varepsilon_{a}\left(\mathbf{k}\right)\right)+\delta\left(\varepsilon_{d}\left(\mathbf{k}'\right)-\varepsilon_{b}\left(\mathbf{k}\right)\right)\right]\mathcal{M}_{cd}^{ab}\left(\mathbf{k},\mathbf{k}'\right)\rho_{cd}\left(\mathbf{k}'\right)\right.\nonumber \\
 & \qquad\left.-\delta\left(\varepsilon_{d}\left(\mathbf{k}'\right)-\varepsilon_{c}\left(\mathbf{k}\right)\right)\left(\mathcal{M}_{dd}^{ac}\left(\mathbf{k},\mathbf{k}'\right)\rho_{cb}\left(\mathbf{k}\right)+\rho_{ac}\left(\mathbf{k}\right)\mathcal{M}_{dd}^{cb}\left(\mathbf{k},\mathbf{k}'\right)\right)\right\} ,\label{eq:scatt_op}
\end{align}
with
\begin{equation}
\mathcal{M}_{cd}^{ab}\left(\mathbf{k},\mathbf{k}'\right)=n_{\mathrm{imp}}\left|V_{\mathbf{k},\mathbf{k}'}\right|^{2}\braket{a,\mathbf{k}|c,\mathbf{k}'}\braket{d,\mathbf{k}'|b,\mathbf{k}},\label{eq:M_mat}
\end{equation}
where $V_{\mathbf{k},\mathbf{k}'}$ is the Fourier transform of the
scattering potential. For the off-diagnonal elements of the scattering
integral, we find
\begin{align}
I_{\mathrm{coll}}\left[\hat{\rho}\right]\left(\mathbf{k}\right)_{+-} & =\frac{\pi}{\hbar}\int\frac{d^{2}k'}{\left(2\pi\right)^{2}}\left\{ \delta\left(\varepsilon_{+}\left(\mathbf{k}'\right)-\varepsilon_{+}\left(\mathbf{k}\right)\right)\mathcal{M}_{++}^{+-}\left(\mathbf{k},\mathbf{k}'\right)\left[\rho_{++}\left(\mathbf{k}'\right)-\rho_{++}\left(\mathbf{k}\right)\right]\right.\nonumber \\
 & \qquad\left[\delta\left(\varepsilon_{+}\left(\mathbf{k}'\right)-\varepsilon_{-}\left(\mathbf{k}\right)\right)\mathcal{M}_{++}^{+-}\left(\mathbf{k},\mathbf{k}'\right)\rho_{++}\left(\mathbf{k}'\right)-\delta\left(\varepsilon_{-}\left(\mathbf{k}'\right)-\varepsilon_{+}\left(\mathbf{k}\right)\right)\mathcal{M}_{--}^{+-}\left(\mathbf{k},\mathbf{k}'\right)\rho_{++}\left(\mathbf{k}\right)\right]\nonumber \\
 & \qquad\left[\delta\left(\varepsilon_{-}\left(\mathbf{k}'\right)-\varepsilon_{+}\left(\mathbf{k}\right)\right)\mathcal{M}_{--}^{+-}\left(\mathbf{k},\mathbf{k}'\right)\rho_{--}\left(\mathbf{k}'\right)-\delta\left(\varepsilon_{+}\left(\mathbf{k}'\right)-\varepsilon_{-}\left(\mathbf{k}\right)\right)\mathcal{M}_{++}^{+-}\left(\mathbf{k},\mathbf{k}'\right)\rho_{--}\left(\mathbf{k}\right)\right]\nonumber \\
 & \qquad\left.+\delta\left(\varepsilon_{-}\left(\mathbf{k}'\right)-\varepsilon_{-}\left(\mathbf{k}\right)\right)\mathcal{M}_{--}^{+-}\left(\mathbf{k},\mathbf{k}'\right)\left[\rho_{--}\left(\mathbf{k}'\right)-\rho_{--}\left(\mathbf{k}\right)\right]\right\} ,\label{eq:+-_scatt_m_arb}
\end{align}
 where we neglected contributions from the off-diagonal density matrix
elements, which are suppressed by the inverse band distance $1/\Delta\varepsilon_{\mathbf{k}}$.
Form Eq. (\ref{eq:M_mat}), it is easy to see that $\mathcal{M}_{++}^{+-}\left(\mathbf{k},\mathbf{k}'\right)=\mathcal{M}_{++}^{-+}\left(\mathbf{k},\mathbf{k}'\right)$,
as well as $\mathcal{M}_{--}^{+-}\left(\mathbf{k},\mathbf{k}'\right)=\mathcal{M}_{--}^{-+}\left(\mathbf{k},\mathbf{k}'\right)$,
and therefore
\begin{equation}
\tau_{\left(m\right),+-|bb}^{-1}=\tau_{\left(m\right),-+|bb}^{-1},
\end{equation}
as stated in the main text (Eq. \eqref{hermiticity_on_relax_times}). 

An assumption used in the main text was that for the $m=0$ scattering times holds
$\tau_{\left(0\right)a\bar{a}|a\bar{a}}^{-1}=\tau_{\left(0\right)a\bar{a}|\bar{a}a}^{-1}$.
For the $m=0$ channel, we find from Eq. (\ref{eq:scatt_op}):
\begin{align}
\nonumber{}
I_{\mathrm{coll}}\left[\hat{g}_{\left(0\right)}\right]\left(\mathbf{k}\right)_{+-} & =\frac{\pi}{\hbar}\int\frac{d^{2}k'}{\left(2\pi\right)^{2}}\left\{ -\left[\delta\left(\varepsilon_{-}\left(\mathbf{k}'\right)-\varepsilon_{+}\left(\mathbf{k}\right)\right)+\delta\left(\varepsilon_{+}\left(\mathbf{k}'\right)-\varepsilon_{-}\left(\mathbf{k}\right)\right)\right]\mathcal{M}_{--}^{++}\left(\mathbf{k},\mathbf{k}'\right)g_{\left(0\right)+-}\left(k\right)\right.\\
 & \qquad\left.+\left[\delta\left(\varepsilon_{-}\left(\mathbf{k}'\right)-\varepsilon_{+}\left(\mathbf{k}\right)\right)+\delta\left(\varepsilon_{+}\left(\mathbf{k}'\right)-\varepsilon_{-}\left(\mathbf{k}\right)\right)\right]\mathcal{M}_{-+}^{+-}\left(\mathbf{k},\mathbf{k}'\right)g_{\left(0\right)-+}\left(k'\right)\right\} .
\end{align}
We used $g_{\left(0\right),aa}=0$, $\mathcal{M}_{++}^{++}\left(\mathbf{k},\mathbf{k}'\right)=\mathcal{M}_{+-}^{+-}\left(\mathbf{k},\mathbf{k}'\right)$,
$\mathcal{M}_{+-}^{+-}\left(\mathbf{k},\mathbf{k}'\right)=\mathcal{M}_{--}^{--}\left(\mathbf{k},\mathbf{k}'\right)$,
and finally $\mathcal{M}_{--}^{++}\left(\mathbf{k},\mathbf{k}'\right)=\mathcal{M}_{++}^{--}\left(\mathbf{k},\mathbf{k}'\right)$.
Since $\mathcal{M}_{--}^{++}\left(\mathbf{k},\mathbf{k}'\right)=-\mathcal{M}_{-+}^{+-}\left(\mathbf{k},\mathbf{k}'\right)$,
conclude that $\tau_{\left(0\right)+-|+-}^{-1}=\tau_{\left(0\right)+-|-+}^{-1}$.
By symmetry, this extends to $\tau_{\left(0\right)a\bar{a}|a\bar{a}}^{-1}=\tau_{\left(0\right)a\bar{a}|\bar{a}a}^{-1}$.

Importantly, we note that if the Fermi-surface lies in the upper band only, scattering
in the $m=0$ channel becomes ineffective:
\begin{equation}
    \tau_{\left(0\right)ab|cd}^{-1}=0
\end{equation}
Other scattering mechanisms
will then dominate over impurity scattering.

\subsection{Scattering times and symmetries} \label{subsec:scattering_times_symmetries}

Let us first discuss the conditions under which scattering rates have to be real numbers.
Preservation of the Hermitian nature of the density matrix implies that
$\tau_{\left(m\right)aa|bb}$ and $\bar{\tau}_{\left(m\right)aa|bb}$ are always real,  since these rates control the time evolution of diagonal elements of the density matrix, that
correspond to populations in the two different bands. Further constraints arise from combining rotational
and time-reversal (TRS) symmetries. At the microscopic level, TRS implies that the probability amplitude
for a wave-packet centered on state $|b,\mathbf{k'}\rangle$ to be scattered by an impurity into state
$|a,\mathbf{k}\rangle$ is the same as the scattering amplitude from state $|a,\mathbf{-k}\rangle$ to state
$|b,\mathbf{-k'}\rangle$. Combining this with rotational invariance implies that:

\begin{eqnarray}
\tau_{\left(m\right)ab|cd} & = & \tau_{\left(m\right)cd|ab} \nonumber\\
\bar{\tau}_{\left(m\right)ab|cd} & = & - \bar{\tau}_{\left(m\right)cd|ab}
\end{eqnarray}
Explicitely, we deduce the following constraints:
\begin{eqnarray}
\tau_{\left(m\right)aa|\bar{a}\bar{a}} & = & \tau_{\left(m\right)\bar{a}\bar{a}|aa} \in \mathbb{R}\nonumber\\
\bar{\tau}_{\left(m\right) aa|\bar{a}\bar{a}} & = & \bar{\tau}_{\left(m\right) \bar{a}\bar{a}|aa} \in \mathbb{R}\nonumber\\
\tau_{\left(m\right)a\bar{a}|\bar{a}a} & = & \tau_{\left(m\right)\bar{a}a|a\bar{a}} \in \mathbb{R} \nonumber\\
\bar{\tau}_{\left(m\right)a\bar{a}|\bar{a}a} & = & -\bar{\tau}_{\left(m\right)\bar{a}a|a\bar{a}} \in i \mathbb{R} \nonumber\\
\bar{\tau}_{\left(m\right) cd|cd} & = & 0. 
\end{eqnarray}
We also note that for scattering on scalar impurities, in the Born approximation, all scattering times are real.

Here we explore the consequences of rotational invariance and mirror
symmetry on the collision times. For a rotationally invariant system,
we can, assuming $k$-independent scattering times, write the collision
operator in the band basis as
\begin{equation}
\mathcal{C}\left[\hat{g}_{\left(m\right)}\cos m\varphi_{\mathbf{k}}\right]=-\int d\varphi_{\mathbf{k}'}W\left(\varphi_{\mathbf{k}}-\varphi_{\mathbf{k}'}\right)\left[\hat{g}_{\left(m\right)}\cos m\varphi_{\mathbf{k}'}\right].\label{eq:relax_time_as_ang_integral}
\end{equation}
Notice that $W$ is a tensor with components $W_{ab|cd}$ and $\hat{g}_{\left(m\right)}$
is a matrix with components $\hat{g}_{\left(m\right),ab}$. Sums over
the components are implied. We can expand
\begin{equation}
W\left(\varphi_{\mathbf{k}}-\varphi_{\mathbf{k}'}\right)=\sum_{m}w_{\left(m\right)}\cos\left[m\left(\varphi_{\mathbf{k}}-\varphi_{\mathbf{k}'}\right)\right]+\bar{w}_{\left(m\right)}\sin\left[m\left(\varphi_{\mathbf{k}}-\varphi_{\mathbf{k}'}\right)\right],\label{eq:relax_kernel_expand}
\end{equation}
giving
\begin{equation}
\mathcal{C}\left[\hat{g}_{\left(m\right)}\cos m\varphi_{\mathbf{k}}\right]=-\tau_{\left(m\right)}^{-1}\hat{g}_{\left(m\right)}\cos m\varphi_{\mathbf{k}}-\bar{\tau}_{\left(m\right)}^{-1}\hat{g}_{\left(m\right)}\sin m\varphi_{\mathbf{k}},
\end{equation}
such that $\tau_{\left(m\right)}^{-1}=\pi w_{\left(m\right)}$ and
$\bar{\tau}_{\left(m\right)}^{-1}=\pi\bar{w}_{\left(m\right)}$. We
then also have
\begin{equation}
\mathcal{C}\left[\hat{\bar{g}}_{\left(m\right)}\sin m\varphi_{\mathbf{k}}\right]=-\tau_{\left(m\right)}^{-1}\hat{\bar{g}}_{\left(m\right)}\sin m\varphi_{\mathbf{k}}+\bar{\tau}_{\left(m\right)}^{-1}\hat{\bar{g}}_{\left(m\right)}\cos m\varphi_{\mathbf{k}}.
\end{equation}

Next, we consider the mirror symmetry. For concreteness we assume
the system to be symmetric with respect to mirroring across the $x$-axis.
The mirror operation acts on momenta ($\varphi_{\mathbf{k}}\rightarrow-\varphi_{\mathbf{k}}$),
and orbital degrees of freedom: $p_{x}\rightarrow p_{x}$, $p_{y}\rightarrow-p_{y}$.
Overall, we can write it's action on the Hamiltonian as
\begin{equation}
\hat{H}\left(k,\varphi_{\mathbf{k}}\right)\rightarrow\hat{U}\hat{H}\left(k,-\varphi_{\mathbf{k}}\right)\hat{U}^{\dagger},
\end{equation}
where 
\begin{equation}
\hat{U}=\left[\begin{array}{cc}
1 & 0\\
0 & -1
\end{array}\right]=\hat{\sigma}_{z},
\end{equation}
is responsible for flipping the sign in the orbital basis. The eigenstates
of $\hat{H}=d_{x}\hat{\sigma}_{x}+d_{z}\hat{\sigma}_{z}$ read
\begin{align}
\ket{+,\mathbf{k}} & =\cos\left(\Theta_{\mathbf{k}}/2\right)\ket{p_{x}}+\sin\left(\Theta_{\mathbf{k}}/2\right)\ket{p_{y}}\nonumber \\
\ket{-,\mathbf{k}} & =\sin\left(\Theta_{\mathbf{k}}/2\right)\ket{p_{x}}-\cos\left(\Theta_{\mathbf{k}}/2\right)\ket{p_{y}}
\end{align}
with 
\begin{equation}
\Theta_{\mathbf{k}}=\arctan\left(d_{x}/d_{z}\right).
\end{equation}
Under mirroring, the eigenstates transform as
\begin{align*}
\ket{+,\mathbf{k}} & \rightarrow\ket{+,\mathbf{k}}\\
\ket{-,\mathbf{k}} & \rightarrow-\ket{-,\mathbf{k}},
\end{align*}
thus, the mirroring operation is still given by the action of $\hat{U}$
and $\varphi_{\mathbf{k}}\rightarrow-\varphi_{\mathbf{k}}$ (acting
on components of the new basis). 

Let us investigate the action of the mirroring operation $\hat{P}\hat{U}\hat{P}^{\dagger}$,
$\varphi_{\mathbf{k}}\rightarrow-\varphi_{\mathbf{k}}$ on the Berry
connection:
\begin{align}
\mathcal{A}_{ab}\hat{\mathbf{e}}_{\varphi_{\mathbf{k}}} & =-\frac{i}{k}\left(\frac{\partial}{\partial\varphi_{\mathbf{k}}}\Theta_{\mathbf{k}}\right)\left(1-\delta_{ab}\right)\epsilon_{ab}\hat{\mathbf{e}}_{\varphi_{\mathbf{k}}},
\end{align}
We find that $\Theta_{\mathbf{k}}\rightarrow\Theta_{\mathbf{k}}$,
$\left(1-\delta_{ab}\right)\epsilon_{ab}\rightarrow-\left(1-\delta_{ab}\right)\epsilon_{ab}$,
and $\hat{\mathbf{e}}_{\varphi_{\mathbf{k}}}\rightarrow-\hat{\mathbf{e}}_{\varphi_{\mathbf{k}}}$,
such that the Berry connection transforms as a vector:
\begin{equation}
\mathcal{A}_{ab}\hat{\mathbf{e}}_{\varphi_{\mathbf{k}}}\rightarrow\mathcal{A}_{ab}\hat{\mathbf{e}}_{-\varphi_{\mathbf{k}}}.
\end{equation}
The resulting OAM currents in $y$-direction are invariant (the velocity
also flips sign). 

Finally, we apply the mirroring operation to the Quantum Boltzmann
equation. For our purposes it is sufficient to consider a spatially
uniform system, in which the density matrix is diagonal in momentum
$\mathbf{k}$. The QBE in the band basis reads
\begin{equation}
\frac{\partial\hat{f}_{\mathbf{k}}}{\partial t}+i\int\frac{d^{2}\mathbf{k}'}{\left(2\pi\right)^{2}}\left[\hat{h}\left(\mathbf{k}\right)\delta\left(\mathbf{k}-\mathbf{k}'\right)+\hat{H}'\left(\mathbf{k}-\mathbf{k}'\right),\hat{f}_{\mathbf{k}'}\right]=0,\label{eq:Form_inv_neumann_before-1}
\end{equation}
where $\hat{H}'$ corresponds to the perturbation that we ultimately
want to describe with the above collision times. Let subdivide the
density matrix in the following way
\begin{equation}
\hat{f}_{\mathbf{k}}=\hat{f}_{\mathbf{k}}^{\mathrm{even}}+\hat{f}_{\mathbf{k}}^{\mathrm{odd}},
\end{equation}
with
\begin{align}
\hat{U}\hat{f}_{k,-\varphi_{\mathbf{k}}}^{\mathrm{even}}\hat{U}^{\dagger} & =\hat{f}_{k,\varphi_{\mathbf{k}}}^{\mathrm{even}}\nonumber \\
\hat{U}\hat{f}_{k,-\varphi_{\mathbf{k}}}^{\mathrm{odd}}\hat{U}^{\dagger} & =-\hat{f}_{k,\varphi_{\mathbf{k}}}^{\mathrm{odd}}.
\end{align}
Since the full Hamiltonian is symmetric under the mirroring operation,
$\hat{H}'\left(\mathbf{k}-\mathbf{k}'\right)$ acting on the density
matrix does not mix the odd and even parts:
\begin{align}
\hat{U}\hat{H}'\hat{f}^{\mathrm{even}}\hat{U}^{\dagger} & =\hat{U}\hat{H}'\hat{U}^{\dagger}\hat{U}\hat{f}^{\mathrm{even}}\hat{U}^{\dagger}=\hat{H}'\hat{f}^{\mathrm{even}}\nonumber \\
\hat{U}\hat{H}'\hat{f}^{\mathrm{odd}}\hat{U}^{\dagger} & =\hat{U}\hat{H}'\hat{U}^{\dagger}\hat{U}\hat{f}^{\mathrm{odd}}\hat{U}^{\dagger}=-\hat{H}'\hat{f}^{\mathrm{odd}},
\end{align}
where we suppressed momentum indices (keeping in mind that the operation
still $\varphi_{\mathbf{k}}\rightarrow-\varphi_{\mathbf{k}}$ has
to be performed). Using Eq. (\ref{eq:relax_time_as_ang_integral}),
we approximate
\begin{equation}
\mathcal{C}\left[\hat{f}\right]_{ab}\left(\mathbf{k}\right)=i\int\frac{d^{2}\mathbf{k}'}{\left(2\pi\right)^{2}}\left[\hat{H}'\left(\mathbf{k}-\mathbf{k}'\right),\hat{f}_{\mathbf{k}'}\right]_{ab}\approx\sum_{cd}\int d\varphi_{\mathbf{k}'}W_{ab|cd}\left(\varphi_{\mathbf{k}}-\varphi_{\mathbf{k}'}\right)f_{\mathbf{k}',cd}.
\end{equation}
We demand that the Kernel $W\left(\varphi_{\mathbf{k}}-\varphi_{\mathbf{k}'}\right)$
preserves the symmetry of$\hat{f}_{\mathbf{k}'}$ in the same way
as $\hat{H}'$. In the expansion
\begin{equation}
\delta\hat{f}\left(\mathbf{k}\right)=\sum_{m}\left(\hat{g}_{\left(m\right)}\cos m\varphi_{\mathbf{k}}+\hat{\bar{g}}_{\left(m\right)}\sin m\varphi_{\mathbf{k}}\right),
\end{equation}
the first term is even under $\varphi_{\mathbf{k}}\rightarrow-\varphi_{\mathbf{k}}$,
while the second is odd. Dividing $\hat{g}_{\left(m\right)}$ and
$\hat{\bar{g}}_{\left(m\right)}$ in diagonal and off-diagonal components,
we see that the individual terms have the following behavior under
the combined action of $\hat{U}$ and $\varphi_{\mathbf{k}}\rightarrow-\varphi_{\mathbf{k}}$:
\begin{align}
\hat{g}_{\left(m\right),aa}\cos m\varphi_{\mathbf{k}} & \rightarrow\hat{g}_{\left(m\right),aa}\cos m\varphi_{\mathbf{k}}\nonumber \\
\hat{g}_{\left(m\right),a\bar{a}}\cos m\varphi_{\mathbf{k}} & \rightarrow-\hat{g}_{\left(m\right),a\bar{a}}\cos m\varphi_{\mathbf{k}}\nonumber \\
\hat{\bar{g}}_{\left(m\right),aa}\sin m\varphi_{\mathbf{k}} & \rightarrow-\hat{\bar{g}}_{\left(m\right),aa}\sin m\varphi_{\mathbf{k}}\nonumber \\
\hat{\bar{g}}_{\left(m\right),a\bar{a}}\sin m\varphi_{\mathbf{k}} & \rightarrow\hat{\bar{g}}_{\left(m\right),a\bar{a}}\sin m\varphi_{\mathbf{k}}
\end{align}
To avoid mixing between terms with different transformation properties,
we conclude that the following relaxation rates (as well as the rates
of reversed processes) must vanish
\begin{align}
\tau_{\left(m\right)aa|a\bar{a}}^{-1} & =0\nonumber \\
\bar{\tau}_{\left(m\right)aa|aa}^{-1} & =0\nonumber \\
\tau_{\left(m\right)aa|\bar{a}a}^{-1} & =0\nonumber \\
\bar{\tau}_{\left(m\right)aa|\bar{a}\bar{a}}^{-1} & =0\nonumber \\
\tau_{\left(m\right)\bar{a}\bar{a}|a\bar{a}}^{-1} & =0\nonumber \\
\bar{\tau}_{\left(m\right)\bar{a}\bar{a}|aa}^{-1} & =0\nonumber \\
\tau_{\left(m\right)\bar{a}\bar{a}|\bar{a}a}^{-1} & =0\nonumber \\
\bar{\tau}_{\left(m\right)\bar{a}\bar{a}|\bar{a}\bar{a}}^{-1} & =0.\label{eq:times_vanish}
\end{align}

Returning to the question of the real nature of scattering rates, the simultaneous
presence of rotational, time-reversal and mirror symmetries garanties that they are all real,
except possibly $\tau_{\left(m\right),a\bar{a}|a\bar{a}}$ and $\bar{\tau}_{\left(m\right),a\bar{a}|\bar{a}a}$.

\subsection{Full set of coupled equations for $m<2$}\label{subsec:full_set}

Using the relations

\begin{align*}
\sin\varphi_{\mathbf{k}}\left(\hat{g}_{\left(2\right)}\cos2\varphi_{\mathbf{k}}+\hat{\bar{g}}_{\left(2\right)}\sin2\varphi_{\mathbf{k}}\right) & \approx-\hat{g}_{\left(2\right)}\frac{\sin\varphi_{\mathbf{k}}}{2}+\hat{\bar{g}}_{\left(2\right)}\frac{\cos\varphi_{\mathbf{k}}}{2}\\
\sin\varphi_{\mathbf{k}}\left(\hat{g}_{\left(1\right)}\cos\varphi_{\mathbf{k}}+\hat{\bar{g}}_{\left(1\right)}\sin\varphi_{\mathbf{k}}\right) & =\hat{g}_{\left(1\right)}\frac{\sin2\varphi_{\mathbf{k}}}{2}+\hat{\bar{g}}_{\left(1\right)}\left(\frac{1}{2}-\frac{\cos2\varphi_{\mathbf{k}}}{2}\right)
\end{align*}
\begin{align*}
\cos\varphi_{\mathbf{k}}\left(\hat{g}_{\left(2\right)}\cos2\varphi_{\mathbf{k}}+\hat{\bar{g}}_{\left(2\right)}\sin2\varphi_{\mathbf{k}}\right) & \approx\hat{g}_{\left(2\right)}\frac{\cos\varphi_{\mathbf{k}}}{2}+\hat{\bar{g}}_{\left(2\right)}\frac{\sin\varphi_{\mathbf{k}}}{2}\\
\cos\varphi_{\mathbf{k}}\left(\hat{g}_{\left(1\right)}\cos\varphi_{\mathbf{k}}+\hat{\bar{g}}_{\left(1\right)}\sin\varphi_{\mathbf{k}}\right) & =\hat{g}_{\left(1\right)}\left(\frac{1}{2}+\frac{\cos2\varphi_{\mathbf{k}}}{2}\right)+\hat{\bar{g}}_{\left(1\right)}\frac{\sin2\varphi_{\mathbf{k}}}{2},
\end{align*}
where we ignored terms with $m>2$, we find a set of equations for
the elements $\hat{g}_{\left(m\right)}$ and $\hat{\bar{g}}_{\left(m\right)}$.
Writing the velocity operator as $\hat{v}=\hat{v}_{\mathrm{d}}+\hat{v}_{\mathrm{od}}$,
where $v_{d}\sim\hat{\mathbf{e}}_{\mathbf{k}}$ and $\hat{v}_{\mathrm{od}}\sim\hat{\mathbf{e}}_{\varphi_{\mathbf{k}}}$.
For $m=1$, we obtain:
\begin{align}
 & \frac{i}{\hbar}\left[\hat{h},\hat{g}_{\left(1\right)}\right]+\frac{iq}{4}\left\{ \hat{v}_{\mathrm{d}},\hat{\bar{g}}_{\left(2\right)}\right\} +\frac{iq}{4}\left\{ \hat{v}_{\mathrm{od}},\hat{g}_{\left(2\right)}\right\} +\frac{iq}{2}\left\{ \hat{v}_{\mathrm{od}},\hat{g}_{\left(0\right)}\right\} +\frac{e}{\hbar}E\frac{\partial\varepsilon_{a}}{\partial k}\frac{\partial f_{0}\left(\varepsilon_{a}\right)}{\partial\varepsilon_{a}}\delta_{ab}\label{eq:boltzmann_components1}\\
 & \qquad=-\tau_{\left(1\right)}^{-1}\hat{g}_{\left(1\right)}+\bar{\tau}_{\left(1\right)}^{-1}\hat{\bar{g}}_{\left(1\right)}\nonumber \\
 & \frac{i}{\hbar}\left[\hat{h},\hat{\bar{g}}_{\left(1\right)}\right]-\frac{iq}{4}\left\{ \hat{v}_{\mathrm{d}},\hat{g}_{\left(2\right)}\right\} +\frac{iq}{4}\left\{ \hat{v}_{\mathrm{od}},\hat{\bar{g}}_{\left(2\right)}\right\} +\frac{iq}{2}\left\{ \hat{v}_{\mathrm{d}},\hat{g}_{\left(0\right)}\right\} -i\frac{e}{\hbar}E\hat{\mathcal{A}}\left[f_{0}\left(\varepsilon_{m}\right)-f_{0}\left(\varepsilon_{n}\right)\right]\label{eq:boltzmann_components2}\\
 & \qquad=-\tau_{\left(1\right)}^{-1}\hat{\bar{g}}_{\left(1\right)}-\bar{\tau}_{\left(1\right)}^{-1}\hat{g}_{\left(1\right)}.\nonumber 
\end{align}
Similarly, for the $m=2$ equations, we find:
\begin{align}
\frac{i}{\hbar}\left[\hat{h},\hat{g}_{\left(2\right)}\right]-\frac{iq}{4}\left\{ \hat{v}_{\mathrm{d}},\hat{\bar{g}}_{\left(1\right)}\right\} +\frac{iq}{4}\left\{ \hat{v}_{\mathrm{od}},\hat{g}_{\left(1\right)}\right\}  & =-\tau_{\left(2\right)}^{-1}\hat{g}_{\left(2\right)}+\bar{\tau}_{\left(2\right)}^{-1}\hat{\bar{g}}_{\left(2\right)}\label{eq:boltzmann_components3}\\
\frac{i}{\hbar}\left[\hat{h},\hat{\bar{g}}_{\left(2\right)}\right]+\frac{iq}{4}\left\{ \hat{v}_{\mathrm{d}},\hat{g}_{\left(1\right)}\right\} +\frac{iq}{4}\left\{ \hat{v}_{\mathrm{od}},\hat{\bar{g}}_{\left(1\right)}\right\}  & =-\tau_{\left(2\right)}^{-1}\hat{\bar{g}}_{\left(2\right)}-\bar{\tau}_{\left(2\right)}^{-1}\hat{g}_{\left(2\right)}.\label{eq:boltzmann_components4}
\end{align}
Additionally, we have the $m=0$ component:
\begin{equation}
\frac{i}{\hbar}\left[\hat{h},\hat{g}_{\left(0\right)}\right]+\frac{iq}{4}\left\{ \hat{v}_{\mathrm{d}},\hat{\bar{g}}_{\left(1\right)}\right\} +\frac{iq}{4}\left\{ \hat{v}_{\mathrm{od}},\hat{g}_{\left(1\right)}\right\} =-\tau_{\left(0\right)}^{-1}\hat{g}_{\left(0\right)}.\label{eq:boltzmann_components5}
\end{equation}
It is useful to divide the equations (\ref{eq:boltzmann_components1})-(\ref{eq:boltzmann_components5})
into diagonal and off-diagonal parts. One motivation for this is that the large energy scale
$\Delta \epsilon$ appears only in off-diagonal parts. Another is that the off-diagonal elements enter expectation values for OAM currents and densities while diagonal elements describe charge flow.

We begin with Eq. (\ref{eq:boltzmann_components5}).
For the diagonal components of the $m=0$ sector, we find: 
\begin{equation}
\frac{iq}{2}v_{aa}\bar{g}_{\left(1\right),aa}+\frac{iq}{4}\left(v_{\bar{a}a}g_{\left(1\right),a\bar{a}}+v_{a\bar{a}}g_{\left(1\right),\bar{a}a}\right)=-\sum_{cd}\tau_{\left(0\right),aa|cd}^{-1}g_{\left(0\right),cd}.\label{eq:m_0_diag}
\end{equation}
Note that charge conservation further restricts the scattering times in this channel.
For the off-diagonal part of the $m=0$ sector, we obtain:
\begin{equation}
\frac{ai}{\hbar}\Delta\varepsilon g_{\left(0\right),a\bar{a}}+\frac{iq}{4}\left(v_{aa}+v_{\bar{a}\bar{a}}\right)\bar{g}_{\left(1\right),a\bar{a}}+\frac{iq}{4}v_{a\bar{a}}\left(g_{\left(1\right),aa}+g_{\left(1\right),\bar{a}\bar{a}}\right)=-\sum_{cd}\tau_{\left(0\right),a\bar{a}|cd}^{-1}g_{\left(0\right),cd}.\label{eq:m_0_off}
\end{equation}
The diagonal and off-diagonal elements of the $\cos\varphi_{\mathbf{k}}$
projection for $m=1$ {[}Eq. (\ref{eq:boltzmann_components1}){]}
read:
\begin{align}
 & \frac{iq}{2}v_{aa}\bar{g}_{\left(2\right),aa}+\frac{iq}{4}\left(v_{\bar{a}a}g_{\left(2\right),a\bar{a}}+v_{a\bar{a}}g_{\left(2\right),\bar{a}a}\right)+\frac{iq}{2}\left(v_{\bar{a}a}g_{\left(0\right),a\bar{a}}+v_{a\bar{a}}g_{\left(0\right),\bar{a}a}\right)+\frac{e}{\hbar}E\frac{\partial\varepsilon_{a}}{\partial k}\frac{\partial f_{0}\left(\varepsilon_{a}\right)}{\partial\varepsilon_{a}}\delta_{ab}\label{eq:m_1_cos_diag}\\
 & \quad=-\sum_{cd}\left(\tau_{\left(1\right),aa|cd}^{-1}g_{\left(1\right),cd}-\bar{\tau}_{\left(1\right),aa|cd}^{-1}\bar{g}_{\left(1\right),cd}\right)
\end{align}
\begin{align}
 & \frac{ai}{\hbar}\Delta\varepsilon g_{\left(1\right),a\bar{a}}+\frac{iq}{4}\left(v_{aa}+v_{\bar{a}\bar{a}}\right)\bar{g}_{\left(2\right),\bar{a}a}+\frac{iq}{4}v_{a\bar{a}}\left(g_{\left(2\right),aa}+g_{\left(2\right),\bar{a}\bar{a}}\right)+\frac{iq}{4}v_{a\bar{a}}\left(g_{\left(0\right),aa}+g_{\left(0\right),\bar{a}\bar{a}}\right)\label{eq:m_1_cos_off}\\
 & \quad=-\sum_{cd}\left(\tau_{\left(1\right),a\bar{a}|cd}^{-1}g_{\left(1\right),cd}-\bar{\tau}_{\left(1\right),a\bar{a}|cd}^{-1}\bar{g}_{\left(1\right),cd}\right)
\end{align}
whereas for the $\sin\varphi_{\mathbf{k}}$ projection {[}Eq. (\ref{eq:boltzmann_components2}){]}
we have
\begin{equation}
-\frac{iq}{2}v_{aa}g_{\left(2\right),aa}+\frac{iq}{4}\left(v_{\bar{a}a}\bar{g}_{\left(2\right),a\bar{a}}+v_{a\bar{a}}\bar{g}_{\left(2\right),\bar{a}a}\right)+iqv_{aa}g_{\left(0\right),aa}=-\sum_{cd}\left(\tau_{\left(1\right),aa|cd}^{-1}\bar{g}_{\left(1\right),cd}+\bar{\tau}_{\left(1\right),aa|cd}^{-1}g_{\left(1\right),cd}\right)\label{eq:m_1_sin_diag}
\end{equation}
\begin{align}
 & \frac{ai}{\hbar}\Delta\varepsilon\bar{g}_{\left(1\right),a\bar{a}}-\frac{iq}{4}\left(v_{aa}+v_{\bar{a}\bar{a}}\right)g_{\left(2\right),a\bar{a}}+\frac{iq}{4}v_{a\bar{a}}\left(\bar{g}_{\left(2\right),aa}+\bar{g}_{\left(2\right),\bar{a}\bar{a}}\right)+\frac{iq}{2}\left(v_{aa}+v_{\bar{a}\bar{a}}\right)g_{\left(0\right),a\bar{a}}+i\frac{e}{\hbar}E\mathcal{A}_{a\bar{a}}\left[f_{0}\left(\varepsilon_{a}\right)-f_{0}\left(\varepsilon_{\bar{a}}\right)\right]\label{eq:m_1_sin_off}\\
 & \quad=-\sum_{cd}\left(\tau_{\left(1\right),a\bar{a}|cd}^{-1}\bar{g}_{\left(1\right),cd}+\bar{\tau}_{\left(1\right),a\bar{a}|cd}^{-1}g_{\left(1\right),cd}\right).
\end{align}
We now turn to the $m=2$ sector. Let us start with the diagonal and
off-diagonal components of the $\cos2\varphi_{\mathbf{k}}$ projection
{[}Eq. (\ref{eq:boltzmann_components3}){]}:
\begin{equation}
-\frac{iq}{4}v_{aa}\bar{g}_{\left(1\right),aa}+\frac{iq}{4}\left(v_{\bar{a}a}g_{\left(1\right),a\bar{a}}+v_{a\bar{a}}g_{\left(1\right),\bar{a}a}\right)=-\sum_{cd}\left(\tau_{\left(2\right),aa,cd}^{-1}g_{\left(2\right),cd}-\bar{\tau}_{\left(2\right),aa,cd}^{-1}\bar{g}_{\left(2\right),cd}\right)\label{eq:m_2_cos_diag}
\end{equation}
\begin{equation}
\frac{ai}{\hbar}\Delta\varepsilon g_{\left(2\right),a\bar{a}}-\frac{iq}{4}\left(v_{aa}+v_{\bar{a}\bar{a}}\right)\bar{g}_{\left(1\right),a\bar{a}}+\frac{iq}{4}v_{a\bar{a}}\left(g_{\left(1\right),aa}+g_{\left(1\right),\bar{a}\bar{a}}\right)=-\sum_{cd}\left(\tau_{\left(2\right),a\bar{a},cd}^{-1}g_{\left(2\right),cd}-\bar{\tau}_{\left(2\right),a\bar{a},cd}^{-1}\bar{g}_{\left(2\right),cd}\right).\label{eq:m_2_cos_off}
\end{equation}
Finally, for the $\sin2\varphi_{\mathbf{k}}$ projection {[}Eq. (\ref{eq:boltzmann_components3}){]}
we find:
\begin{equation}
\frac{iq}{4}v_{aa}g_{\left(1\right),aa}+\frac{iq}{4}\left(v_{\bar{a}a}\bar{g}_{\left(1\right),a\bar{a}}+v_{a\bar{a}}\bar{g}_{\left(1\right),\bar{a}a}\right)=-\sum_{cd}\left(\tau_{\left(2\right),aa|cd}^{-1}\bar{g}_{\left(2\right),cd}+\bar{\tau}_{\left(2\right),aa|cd}^{-1}g_{\left(2\right),cd}\right)\label{eq:m_2_sin_diag}
\end{equation}
\begin{equation}
\frac{ai}{\hbar}\Delta\varepsilon\bar{g}_{\left(2\right),a\bar{a}}+\frac{iq}{4}\left(v_{aa}+v_{\bar{a}\bar{a}}\right)g_{\left(1\right),a\bar{a}}+\frac{iq}{4}v_{a\bar{a}}\left(\bar{g}_{\left(1\right),aa}+\bar{g}_{\left(1\right),\bar{a}\bar{a}}\right)=-\sum_{cd}\left(\tau_{\left(2\right),a\bar{a}|cd}^{-1}\bar{g}_{\left(2\right),cd}+\bar{\tau}_{\left(2\right),a\bar{a}|cd}^{-1}g_{\left(2\right),cd}\right).\label{eq:m_2_sin_off}
\end{equation}
Note that many scattering times here vanish due to the results presented in \ref{subsec:scattering_times_symmetries}.

The small parameters
\begin{align*}
\xi & =\frac{\hbar\tau^{-1}}{\Delta\varepsilon}\ll1
\end{align*}
and 
\[
\xi_{q}=\frac{\hbar\tau_{q}^{-1}}{\Delta\varepsilon}\lesssim\xi
\]
establish a hierarchy in Eqs. (\ref{eq:m_0_diag})-(\ref{eq:m_2_sin_off}).
We disentangle the equations by power-counting. First, we note that
all off-diagonal density matrix elements appear in combinations $\sim\Delta\varepsilon g_{\left(m\right),a\bar{a}}$,
$\sim\Delta\varepsilon\bar{g}_{\left(m\right),a\bar{a}}$, while the
diagonal terms do not. When the equations are inverted, since $\Delta\varepsilon\gg\hbar\tau^{-1}$
and $\Delta\varepsilon\gg\hbar\tau^{-1}$, we necessarily find $g_{\left(m\right),a\bar{a}}\sim\Delta\varepsilon^{-1}$,
$\bar{g}_{\left(m\right),a\bar{a}}\sim\Delta\varepsilon^{-1}$. Applying
this observation to Eq. (\ref{eq:m_1_cos_diag}), and noticing that,
following Eq. (\ref{eq:times_vanish}), $\bar{\tau}_{\left(1\right),aa|cc}^{-1}=0$,
we, find, to leading order in $\xi$, $\xi_{q}$,
\begin{equation}
\frac{iq}{2}v_{aa}\bar{g}_{\left(2\right),aa}+\frac{e}{\hbar}E\frac{\partial\varepsilon_{a}}{\partial k}\frac{\partial f_{0}\left(\varepsilon_{a}\right)}{\partial\varepsilon_{a}}\delta_{ab}=-\sum_{b}\tau_{\left(1\right),aa|bb}^{-1}g_{\left(1\right),bb}.\label{eq:simpl_4_1}
\end{equation}
Similarly, Eq. (\ref{eq:m_2_sin_diag}) becomes
\begin{equation}
\frac{iq}{4}v_{aa}g_{\left(1\right),aa}=-\sum_{b}\tau_{\left(2\right),aa|bb}^{-1}\bar{g}_{\left(2\right),bb}.\label{eq:simpl_4_2}
\end{equation}
These equations form a closed set and contribute two of the four governing
equations in the main text (after dropping the inter-band scattering
terms). 

Apart from Eq. (\ref{eq:m_1_cos_diag}), the electric field also appears
in Eq. (\ref{eq:m_1_sin_off}). Applying the above arguments, we disregard
off-diagonal contribution with a prefactor of $\tau_{q}^{-1}$ or
$\tau^{-1}$ in Eq. (\ref{eq:m_1_sin_off}), yielding
\begin{equation}
\frac{ai}{\hbar}\Delta\varepsilon\bar{g}_{\left(1\right),a\bar{a}}+\frac{iq}{4}v_{a\bar{a}}\left(\bar{g}_{\left(2\right),aa}+\bar{g}_{\left(2\right),\bar{a}\bar{a}}\right)+i\frac{e}{\hbar}E\mathcal{A}_{a\bar{a}}\left[f_{0}\left(\varepsilon_{a}\right)-f_{0}\left(\varepsilon_{\bar{a}}\right)\right]=-\sum_{b}\bar{\tau}_{\left(1\right),a\bar{a}|bb}^{-1}g_{\left(1\right),bb}.\label{eq:simpl_4_3}
\end{equation}
The three equations (\ref{eq:simpl_4_1})-(\ref{eq:simpl_4_3}) provide
a closed set, where all density matrix components are of order $\xi^{0}$
and $\xi_{q}^{0}$ (notice that $g_{\left(1\right),aa}\sim\tau$ and
$\bar{g}_{\left(2\right),aa}\sim\tau^{2}\tau_{q}^{-1}$).

Finally, since the experimental quantity of interest -- the density
of OAM -- is encoded in $g_{\left(0\right),\bar{a}a}$, we need to
consider Eq. (\ref{eq:m_0_off}):
\begin{equation}
\frac{ai}{\hbar}\Delta\varepsilon g_{\left(0\right),a\bar{a}}+\frac{iq}{4}\left(v_{aa}+v_{\bar{a}\bar{a}}\right)\bar{g}_{\left(1\right),a\bar{a}}+\frac{iq}{4}v_{a\bar{a}}\left(g_{\left(1\right),aa}+g_{\left(1\right),\bar{a}\bar{a}}\right)=-\sum_{b}\tau_{\left(0\right),a\bar{a}|b\bar{b}}^{-1}g_{\left(0\right),b\bar{b}},\label{eq:simpl_4_4}
\end{equation}
where we used $g_{\left(0\right),aa}=0$ (the charge density decouples
from the electric field in transverse charge flows with $\mathbf{q}\bot\mathbf{E}$).
Since $g_{\left(0\right),a\bar{a}}\sim\xi$, we ignored it's back-coupling
to $\bar{g}_{\left(1\right),a\bar{a}}$ in Eq. (\ref{eq:m_1_sin_off}).
Thus we obtain the four coupled equations given in the main text,
which represent a reduced version of the quantum Boltzmann equation.

As a safety check, let us now reexamine the consistency of Eqs. (\ref{eq:simpl_4_1})-(\ref{eq:simpl_4_4})
with respect to power counting in $\xi$ and $\xi_{q}$. To be more
specific, we introduce a fourth energy scale that quantifies the action
of the electric field:
\[
\frac{e}{\hbar}E\frac{\partial\varepsilon_{a}}{\partial k}\frac{\partial f_{0}\left(\varepsilon_{a}\right)}{\partial\varepsilon_{a}}\sim\tau_{E}^{-1}.
\]
In terms of the four time scales $\tau$ (scattering), $\tau_{q}$
(inhomogeneity), $\tau_{E}$ and $\tau_{\varepsilon}=\hbar\Delta\varepsilon^{-1}$,
we find, schematically, 
\begin{align*}
g_{\left(1\right),aa} & \sim\tau\tau_{E}^{-1}\\
\bar{g}_{\left(2\right),aa}\sim & \tau^{2}\tau_{E}^{-1}\tau_{q}^{-1}\\
\bar{g}_{\left(1\right),a\bar{a}} & \sim\tau_{\varepsilon}\tau_{E}^{-1}\\
g_{\left(0\right),a\bar{a}} & \sim\xi_{q}\bar{g}_{\left(1\right),a\bar{a}}+\xi_{q}\bar{g}_{\left(2\right),aa}+\xi g_{\left(1\right),aa}.
\end{align*}
In Eq. (\ref{eq:m_1_cos_diag}), we ignored density matrix components
$g_{\left(0\right),a\bar{a}}$ and
\[
g_{\left(2\right),a\bar{a}}\sim\xi_{q}g_{\left(1\right),aa}+\xi\bar{g}_{\left(2\right),aa},
\]
which is consistent with our leading order approximation. In Eq. (\ref{eq:m_1_sin_off}),
we, again, ignored $g_{\left(2\right),a\bar{a}}$, $g_{\left(0\right),a\bar{a}}$
and a right hand side contribution of order $\tau^{-1}\bar{g}_{\left(1\right)}\sim\xi\tau_{E}^{-1}$,
again in agreement with our approximation. Finally, in Eq. (\ref{eq:m_2_sin_diag}),
we neglected a terms of the form
\[
\tau_{q}^{-1}\bar{g}_{\left(1\right),a\bar{a}}\sim\xi_{q}\tau_{E}^{-1}
\]
and $\bar{\tau}^{-1}g_{\left(2\right),a\bar{a}}\sim\bar{\tau}^{-1}\xi$.
We conclude that our calculations are consistent and hold to leading
order in $\xi$. As a final caveat, we note that in the limit of extremely
small inhomogeneity $\tau_{q}^{-1}\ll\tau^{-1}$, the condition $\xi\ll\tau_{q}^{-1}/\tau^{-1}$
might become important for our theory to hold.

Notice that the equations for $g_{(0),aa}$ (diagonal elements), $g_{(1),a\bar{a}}$ (off- diagonal), $\bar{g}_{(1),aa}$ ( diagonal),  $g_{(2),ab}$ (diagonal and off-diagonal), $\bar{g}_{(2),a\bar{a}}$ (off-diagonal), fully decouple from the equations governing the OAM flow (Eqs. \eqref{eq:g_(0)_Eq_final}-\eqref{eq:g_bar_(2)_Eq_final} of the main text), to leading order in the small parameters and considering the restrictions of Eqs. \eqref{eq:times_vanish}.

\subsection{Including diagonal inter-band scattering\label{subsec:inter-band-times}}

We want to include scatterings of the type $\tau_{\left(m\right)aa|bb}^{-1}$. This will modify the right hand sides of Eqs. \eqref{eq:eq:g_(1)_Eq_final}, \eqref{eq:g_bar_(2)_Eq_final}. Written in components, the two equations become:
We then have from the second and fourth equations above
\begin{align}
\frac{iq}{2}v_{+}\bar{g}_{\left(2\right),++}+eEv_{+,\mathbf{k}}\frac{\partial f_{0}\left(\varepsilon_{+,\mathbf{k}}\right)}{\partial\varepsilon_{+,\mathbf{k}}} & =-\tau_{\left(1\right),++|++}^{-1}g_{\left(1\right),++}-\tau_{\left(1\right),++|--}^{-1}g_{\left(1\right),--}\\
\frac{iq}{2}v_{-}\bar{g}_{\left(2\right),--}+eEv_{-,\mathbf{k}}\frac{\partial f_{0}\left(\varepsilon_{-,\mathbf{k}}\right)}{\partial\varepsilon_{-,\mathbf{k}}} & =-\tau_{\left(1\right),--|++}^{-1}g_{\left(1\right),++}-\tau_{\left(1\right),--|--}^{-1}g_{\left(1\right),--}
\end{align}
\begin{align}
\frac{iq}{2}v_{+}g_{\left(1\right),++}&=-\tau_{\left(2\right),++|++}^{-1}\bar{g}_{\left(2\right),++}-\tau_{\left(2\right),++|--}^{-1}\bar{g}_{\left(2\right),--}
´\\
\frac{iq}{2}v_{-}g_{\left(1\right),--}&=-\tau_{\left(2\right),--|++}^{-1}\bar{g}_{\left(2\right),++}-\tau_{\left(2\right),--|--}^{-1}\bar{g}_{\left(2\right),--}
\end{align}
From the last two equations, it is clear that we can write 
\begin{align}
\bar{g}_{\left(2\right),++} & =-\frac{iq}{2}v_{+}\tau_{A}g_{\left(1\right),++}-\frac{iq}{2}v_{-}\tau_{B}g_{\left(1\right),--}\\
\bar{g}_{\left(2\right),--} & =-\frac{iq}{2}v_{+}\tau_{C}g_{\left(1\right),++}-\frac{iq}{2}v_{-}\tau_{D}g_{\left(1\right),--}.
\end{align}
It then follows that the first two equations can be recast as
\begin{align}
eEv_{+,\mathbf{k}}\frac{\partial f_{0}\left(\varepsilon_{+,\mathbf{k}}\right)}{\partial\varepsilon_{+,\mathbf{k}}} & =-\left(\tau_{\left(1\right),++|++}^{-1}+\frac{1}{4}q^{2}v_{+}^{2}\tau_{A}\right)g_{\left(1\right),++}-\left(\tau_{\left(1\right),++|--}^{-1}+\frac{1}{4}q^{2}v_{+}v_{-}\tau_{B}\right)g_{\left(1\right),--}\\
eEv_{-,\mathbf{k}}\frac{\partial f_{0}\left(\varepsilon_{-,\mathbf{k}}\right)}{\partial\varepsilon_{-,\mathbf{k}}} & =-\left(\tau_{\left(1\right),--|++}^{-1}+\frac{1}{4}q^{2}v_{+}v_{-}\tau_{C}\right)g_{\left(1\right),++}-\left(\tau_{\left(1\right),--|--}^{-1}+\frac{1}{4}q^{2}v_{-}^{2}\tau_{D}\right)g_{\left(1\right),--}
\end{align}
Inverting these equations is trivial but involves some algebra. For the $+$-band, e.g., we find
\begin{equation}
g_{\left(1\right),++} =\frac{\left[\left(\tau_{\left(1\right),--|--}^{-1}+\frac{1}{4}q^{2}v_{-}^{2}\tau_{D}\right)F_{E,+}-\left(\tau_{\left(1\right),++|--}^{-1}+\frac{1}{4}q^{2}v_{+}v_{-}\tau_{B}\right)F_{E,-}\right]}{\left(\tau_{\left(1\right),++|--}^{-1}+\frac{1}{4}q^{2}v_{+}v_{-}\tau_{B}\right)\left(\tau_{\left(1\right),--|++}^{-1}+\frac{1}{4}q^{2}v_{+}v_{-}\tau_{C}\right)-\left(\tau_{\left(1\right),++|++}^{-1}+\frac{1}{4}q^{2}v_{+}^{2}\tau_{A}\right)\left(\tau_{\left(1\right),--|--}^{-1}+\frac{1}{4}q^{2}v_{-}^{2}\tau_{D}\right)}F_{E,+},
\label{inter_bang_g}
\end{equation}
where we have abbreviated
\begin{align*}
F_{E,\pm} & =eEv_{\pm,\mathbf{k}}\frac{\partial f_{0}\left(\varepsilon_{\pm,\mathbf{k}}\right)}{\partial\varepsilon_{\pm,\mathbf{k}}}.
\end{align*}
This expression reduces to the results of the main text, if inter-band scattering is absent.
Expanding for small $q$, we find that, the basic
structure of Eq. \eqref{inter_bang_g} will be given by 
\begin{equation}
   g_{\left(1\right),++}\approx\frac{1}{\tau_{\mathrm{eff},1,+}^{-1}+\frac{1}{4}q^{2}v_{\mathrm{eff},+}^{2}\tau_{\mathrm{eff},2,+}}F_{E,+}-\frac{1}{\tau_{\mathrm{eff},1,-}^{-1}+\frac{1}{4}q^{2}v_{\mathrm{eff},-}^{2}\tau_{\mathrm{eff},2,-}}F_{E,-}, 
   \label{gpp_long}
\end{equation}
and the quantities marked ``$\mathrm{eff}$'' are composed of the
original scattering times and the velocities in the two bands. This form of $g_{(1),++}$ leads to a conductivity of the form of Eq. \ref{eq:conductivity}, where the velocities and scattering times, again, take their effective values to include inter-band scattering effects.

\subsection{Derivation of the diffusion equation for orbital currents in inhomogeneous
fields\label{subsec:Derivation-of-the_diff}}

In this appendix we derive Eq. (\ref{eq:current_diffusion}) starting from a strip geometry. We first invert Eq. (\ref{eq:Ohm_mom_space})
to write
\begin{equation}
\int_{-w/2}^{w/2} dy'\sigma_{H}^{-1}\left(y-y'\right)J_{L_{z}}\left(y'\right)=E_{x},\label{eq:inverse_Ohm}
\end{equation}
where $w$ is the width of the strip. The kernel $\sigma_{H}^{-1}\left(y-y'\right)$ is found by performing
the summation 

\begin{equation}
\sigma_{H}^{-1}\left(y\right)=\frac{1}{w}\sum_{n=-\infty}^{\infty}\sigma_{H}^{-1}\left(q_{n}\right)e^{iq_{n}y},
\end{equation}
where $q_{n}=2\pi n/w$. Representing the sum over $q_{n}$ as an integral over delta functions
and using the identity $\sum_{n=-\infty}^{\infty}\delta\left(n-\alpha\right)=\sum_{m=-\infty}^{\infty}e^{2\pi im\alpha}$,
we find
\[
\sigma_{H}^{-1}\left(y\right)=\frac{1}{w}\sum_{n=-\infty}^{\infty}\sigma_{H}^{-1}\left(q_{n}\right)e^{iq_{n}y}=\sum_{m=-\infty}^{\infty}\sigma_{H,\mathrm{part.}}^{-1}\left(y+wm\right),
\]
where $\sigma_{H,\mathrm{part.}}^{-1}\left(y\right)=\int_{-\infty}^{\infty}\frac{dq}{2\pi}e^{iqy}\sigma^{-1}\left(q\right)$.
We find
\[
\sigma_{H,\mathrm{part.}}^{-1}\left(y\right)=\frac{\delta\left(y\right)}{\sigma_{H,0}^{\mathrm{int.}}+4\sigma_{H,0}^{\mathrm{n.l.}}}-\frac{\sigma_{H,0}^{\mathrm{ext.}}-4\sigma_{H,0}^{\mathrm{n.l.}}}{\sigma_{H,0}^{\mathrm{int.}}+4\sigma_{H,0}^{\mathrm{n.l.}}}\frac{e^{-\frac{|y|}{\bar{l}}\sqrt{\frac{\sigma_{H,0}^{\mathrm{ext.}}+\sigma_{H,0}^{\mathrm{int.}}}{\sigma_{H,0}^{\mathrm{int.}}+4\sigma_{H,0}^{\mathrm{n.l.}}}}}}{2\bar{l}\sqrt{\left(\sigma_{H,0}^{\mathrm{ext.}}+\sigma_{H,0}^{\mathrm{int.}}\right)\left(\sigma_{H,0}^{\mathrm{int.}}+4\sigma_{H,0}^{\mathrm{n.l.}}\right)}}
\]
with 
\begin{align*}
\bar{l} & =\frac{1}{2}v_{F}\sqrt{\tau_{\left(1\right),+}\tau_{\left(2\right),+}}.
\end{align*}
We introduce
\begin{align*}
\sigma_{1} & =\sigma_{H,0}^{\mathrm{ext.}}+\sigma_{H,0}^{\mathrm{int.}}\\
\sigma_{2} & =\sigma_{H,0}^{\mathrm{int.}}+4\sigma_{H,0}^{\mathrm{n.l.}}\\
\sigma_{3} & =\sigma_{H,0}^{\mathrm{ext.}}-4\sigma_{H,0}^{\mathrm{n.l.}}
\end{align*}
and write
\[
\sigma_{H,\mathrm{part.}}^{-1}\left(y\right)=\frac{\delta\left(y\right)}{\sigma_{2}}-\frac{\sigma_{3}}{\sigma_{2}}\frac{e^{-\frac{|y|}{\bar{l}}\sqrt{\frac{\sigma_{1}}{\sigma_{2}}}}}{2\bar{l}\sqrt{\sigma_{1}\sigma_{2}}}
\]
We are only interested in values $-w/2<y<w/2$, so that $-\left|y+mw\right|=-mw-y$
for $m>0$ and $-\left|y+mw\right|=mw+y$ for $m<0$. We therefore
have
\begin{align*}
\sigma_{H}^{-1}\left(y\right)=\frac{1}{w}\sum_{n=-\infty}^{\infty}\sigma_{H}^{-1}\left(q_{n}\right)e^{iq_{n}y} & =\frac{\delta\left(y\right)}{\sigma_{2}}-\frac{\sigma_{3}/\sigma_{2}}{2\bar{l}\sqrt{\sigma_{1}\sigma_{2}}}\left(e^{-\frac{|y|}{\bar{l}}\sqrt{\frac{\sigma_{1}}{\sigma_{2}}}}+2\cosh\left(\frac{y}{\bar{l}}\sqrt{\frac{\sigma_{1}}{\sigma_{2}}}\right)\frac{1}{e^{\frac{w}{\bar{l}}\sqrt{\frac{\sigma_{1}}{\sigma_{2}}}}-1}\right).
\end{align*}
The current flow is determined by the integral equation
\[
\int dy'\sigma_{H}^{-1}\left(y-y'\right)J_{L_{z}}\left(y'\right)=E_{x},
\]
which we write out to obtain
\[
\frac{J_{L_{z}}\left(y\right)}{\sigma_{2}}-\frac{\sigma_{3}/\sigma_{2}}{2\bar{l}\sqrt{\sigma_{1}\sigma_{2}}}\int dy'\left(e^{-\frac{|y-y'|}{\bar{l}}\sqrt{\frac{\sigma_{1}}{\sigma_{2}}}}+2\cosh\left(\frac{y-y'}{\bar{l}}\sqrt{\frac{\sigma_{1}}{\sigma_{2}}}\right)\frac{1}{e^{\frac{w}{\bar{l}}\sqrt{\frac{\sigma_{1}}{\sigma_{2}}}}-1}\right)J_{L_{z}}\left(y'\right)=E_{x}\left(y\right).
\]
We differentiate twice to find
\begin{align*}
\partial_{y}^{2}e^{-\frac{|y-y'|}{\bar{l}}\sqrt{\frac{\sigma_{1}}{\sigma_{2}}}} & =\frac{1}{\bar{l}^{2}}\frac{\sigma_{1}}{\sigma_{2}}e^{-\frac{|y-y'|}{\bar{l}}\sqrt{\frac{\sigma_{1}}{\sigma_{2}}}}-\frac{2}{\bar{l}}\sqrt{\frac{\sigma_{1}}{\sigma_{2}}}e^{-\frac{|y-y'|}{\bar{l}}\sqrt{\frac{\sigma_{1}}{\sigma_{2}}}}\delta\left(y-y'\right)\\
\partial_{y}^{2}\cosh\left(\frac{y-y'}{\bar{l}}\sqrt{\frac{\sigma_{1}}{\sigma_{2}}}\right) & =\frac{1}{\bar{l}^{2}}\frac{\sigma_{1}}{\sigma_{2}}\cosh\left(\frac{y-y'}{\bar{l}}\sqrt{\frac{\sigma_{1}}{\sigma_{2}}}\right).
\end{align*}
All in all, we have
\begin{align*}
 & \frac{J_{L_{z}}''\left(y\right)}{\sigma_{2}}-\frac{\sigma_{3}/\sigma_{2}}{2\bar{l}\sqrt{\sigma_{1}\sigma_{2}}}\int dy'\Biggr[\frac{1}{\bar{l}^{2}}\frac{\sigma_{1}}{\sigma_{2}}e^{-\frac{|y-y'|}{\bar{l}}\sqrt{\frac{\sigma_{1}}{\sigma_{2}}}}-\frac{2}{\bar{l}}\sqrt{\frac{\sigma_{1}}{\sigma_{2}}}e^{-\frac{|y-y'|}{\bar{l}}\sqrt{\frac{\sigma_{1}}{\sigma_{2}}}}\delta\left(y-y'\right)\\
 & \qquad\qquad+2\frac{1}{\bar{l}^{2}}\frac{\sigma_{1}}{\sigma_{2}}\cosh\left(\frac{y-y'}{\bar{l}}\sqrt{\frac{\sigma_{1}}{\sigma_{2}}}\right)\frac{1}{e^{\frac{w}{\bar{l}}\sqrt{\frac{\sigma_{1}}{\sigma_{2}}}}-1}\Biggr]J_{L_{z}}\left(y'\right)=E_{x}''\left(y\right),
\end{align*}
which ultimately yields Eq. (\ref{eq:current_diffusion}):
\[
J_{L_{z}}''\left(y\right)-\frac{1}{\bar{l}^{2}}J_{L_{z}}\left(y\right)=-\frac{\sigma_{1}}{\bar{l}^{2}}E_{x}\left(y\right)+\sigma_{2}E_{x}''\left(y\right).
\]


\end{document}